\documentclass[twocolumn,tighten,times]{aastex63}
\shorttitle{}
\shortauthors{Schad, Dima \& Anan}
\usepackage{enumitem}          
\usepackage{comment} 

\graphicspath{{./}{figures/}}
\begin{document}

\title{\Large{He I spectropolarimetry of a supersonic coronal \\ downflow within a sunspot umbra}}
\author[0000-0002-7451-9804]{Thomas A. Schad}
\affil{National Solar Observatory, 22 Ohia Ku St., Pukalani, Hawaii 96768, USA}
\author[0000-0002-6003-4646]{Gabriel I. Dima}
\affil{High Altitude Observatory, National Center for Atmospheric Research, P.O. Box 3000, Boulder, CO 80307-3000, USA}
\author[0000-0001-6824-1108]{Tetsu Anan}
\affil{National Solar Observatory, 22 Ohia Ku St., Pukalani, Hawaii 96768, USA}

\begin{abstract}  
We report spectropolarimetric observations of a supersonic downflow impacting the lower atmosphere within a large sunspot umbra.  This work is an extension of \citet{schad2016} using observations acquired in the \ion{He}{1} 10830 \mbox{\AA} triplet by the Facility Infrared Spectropolarimeter.  Downflowing material accelerating along a cooled coronal loop reaches peak speeds near 200 km s$^{-1}$ and exhibits both high speed emission and absorption within the umbra, which we determine to be a consequence of the strong height dependence of the radiatively-controlled source function above the sunspot umbra.  Strong emission profiles close to the rest wavelengths but with long red-shifted tails are also observed at the downflow terminus.  From the polarized spectra, we infer longitudinal magnetic field strengths of ${\sim}2.4$ kG in the core portion of the \ion{He}{1} strong emission, which we believe is the strongest ever reported in this line.  Photospheric field strengths along the same line-of-sight are ${\sim}2.8$ kG as inferred using the \ion{Ca}{1} 10839 \AA{} spectral line.  The temperatures of the highest speed \ion{He}{1} absorption and the near rest emission are similar (${\sim}$10 kK), while a differential emission measure analysis using SDO/AIA data indicates significant increases in radiative cooling for temperatures between $\sim$0.5 and 1 MK plasma associated with the downflow terminus.  Combined we interpret these observations in the context of a strong radiative shock induced by the supersonic downflow impacting the low sunspot atmosphere. 
\end{abstract}

\keywords{\href{http://astrothesaurus.org/uat/1503}{Solar Magnetic Fields (1503)}; \href{ http://astrothesaurus.org/uat/1479}{Solar Chromosphere (1479)};  \href{http://astrothesaurus.org/uat/1532}{Solar Transition Region (1532)}; 
\href{http://astrothesaurus.org/uat/1653}{Sunspots (1653)};  \href{http://astrothesaurus.org/uat/1973}{Spectropolarimetry (1973)}}

\section{Introduction} \label{sec:intro} 

Sunspot umbrae, characterized by strong vertical magnetic fields, are intricately structured and vigorously dynamic in their upper layers. In the photosphere, small-scale features include magneto-convective umbral dots and light bridges \citep{solanki2003}. Photospheric velocity fluctuations ($\Delta v{\sim}0.5$ km s$^{-1}$) have peak power near 5 minute periods \citep{khomenko2015}, while waves propagating upward steepen into shocks ($\Delta v{\sim}15$ km s$^{-1}$) that drive 3 minute period chromospheric umbral flashes \citep{beckers1969, rouppe2003}, which exist alongside small-scale brightenings and dynamic fibrils \citep{henriques2020}.  Meanwhile, downward energy deposition from the corona can also generate small brightenings in the umbral transition region (TR) and chromosphere.  These complexities, along with the non-LTE formation of chromospheric diagnostics, challenge the modeling of the umbral atmosphere \citep[see, e.g.,][]{iwai2017ApJ, loukitcheva2017}.

Our focus is on the upper atmospheric drivers of umbral fine structure.  In a study of 60 data-sets from the \textit{Interface Region Imaging Spectrograph} \citep[IRIS:][]{depontieu2014}, \citet{samanta2018} identified 20 cases of supersonic downflows (SSDs) in the \ion{Si}{4} 1403 \AA{} line located within sunspot umbrae. \citet{kleint2014} originally reported the association of SSDs with bright dot features in the TR above sunspots. Their primary spectral signatures include a near stationary component and a strongly redshifted (50-200 km s$^{-1}$) satellite \citep{dere1982, nicolas1982, kjeldseth1988, brynildsen2004, kleint2014,tian2014, nelson2020, nelson2020b}. Some SSDs show a chromospheric redshifted counterpart observable in the \ion{Mg}{2} h and k and \ion{C}{2} lines.  Another class of SSD spectral profile exhibits strong red line asymmetries, but without a clear separation between stationary and shifted components \citep{kjeldseth1988, kleint2014, ishikawa2020}.

\begin{figure*}
    \centering
    \includegraphics[width=0.95\textwidth]{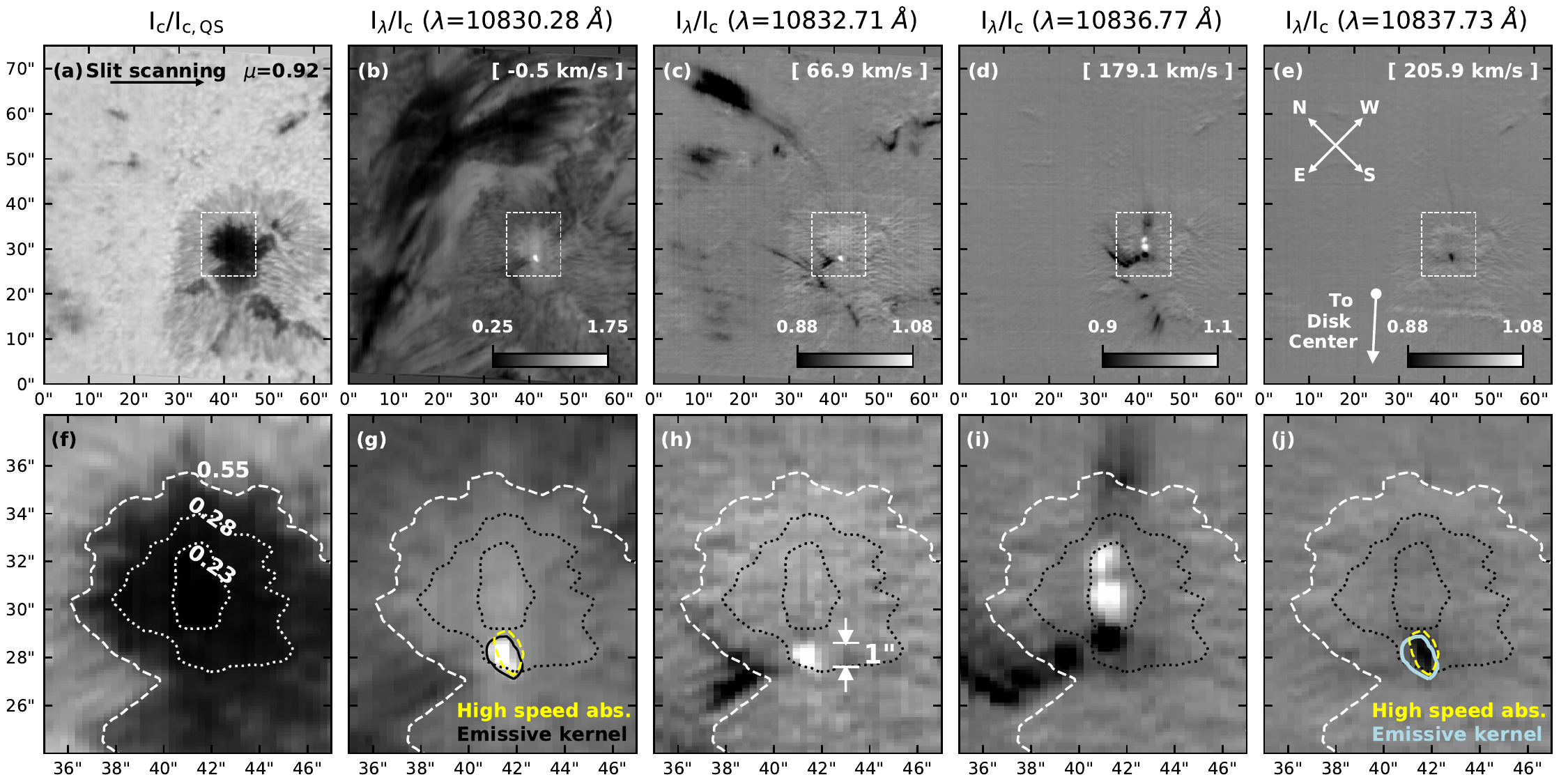}
    \caption{DST/FIRS spectroscopic observations near \ion{He}{1} 10830 \mbox{\AA} of a supersonic downflow draining into a sunspot umbra near N23W11 on 19 September 2011. Panels (a-e) display the full slit-scanned field of view, while panels (f-j) zoom in on the umbral region outlined by a white dashed rectangular in the upper row.  The left-most column shows the continuum intensity normalized to the local quiet sun (see contours in panel (f)).  The other columns contain normalized monochromatic images at progressively larger redshifts from the rest wavelength of the \ion{He}{1} triplet at 10830.3 \mbox{\AA}. These data have been spatially deconvolved to treat instrumental straylight. Dashed yellow and solid (black or blue) contours in (g) and (j) surround the region of high speed absorption and the emissive kernel at rest wavelengths, respectively.} 
    \label{fig:overview}
\end{figure*}

The origin of these SSDs has been suggested by numerous authors to be coronal rain \citep{antolin2012ondisk, ahn2014} or siphon flows along coronal loops \citep{cargill1980, orlando1995}.  Electron densities in the downflowing component observed by \citet{samanta2018} ($\log n_{e} \approx 10.17 \pm 0.69$) are an order of magnitude less dense than the stationary component, supporting a coronal origin.  The short-lived bursty behavior of some downflow profiles (${\sim}$20 seconds in \citet{kleint2014} over a duration of 2 hours) has been linked to the clumpy character of coronal rain \citep{ishikawa2020}.  Meanwhile, the steady nature of the downflows over 10s of minutes observed by \citet{straus2015} and \citet{chitta2016} require a large mass supply, and may support a siphon flow scenario. 

SSDs eventually decelerate and impart their energy into the lower atmosphere, a process important also in solar and stellar flares \citep{fuhrmeister2005, ayres2010, xu2016, graham2020} and eruption downflows \citep{gilbert2013}. Few detailed studies of these transitions have been reported \citep{chitta2016,straus2015}, and there are many open questions, including how deep into the atmosphere this energy can penetrate within the umbra.

In this article, we study an umbral SSD with peak speeds near 200 km s$^{-1}$ observed using spectropolarimetry of the neutral helium 10830 \AA{} triplet, which under equilibrium conditions is formed at temperatures of ${\lesssim}10^{4.5}$ K.  SSDs ($\gtrsim15$ km s$^{-1}$) have previously been observed in \ion{He}{1} 10830 \AA{} by \citet{schmidt2000}, \citet{aznar_cuadrado2007}, and \citet{lagg2007}, particularly near pores, but not to our knowledge within sunspot umbrae, and not at the high speeds studied here.  Umbral \ion{He}{1} typically presents weak absorption spectra ($\lesssim$10\%) modulated by chromospheric oscillations; though, line shapes can be complex \citep{lites1986, centeno2005, anan2018}.  With sensitivity to the Zeeman effect and atomic level polarization \citep{trujillo_bueno2007}, the \ion{He}{1} triplet has been used for sunspot magnetic field measurements \citep{orozco_suarez2005,schad2013, schad2015, joshi2017}. It also is a good diagnostic of coronal rain off-limb \citep{schad2018}. The observations presented here are associated with cool draining filament material, first studied in \citet{schad2016}, that originates from a highly suspended ($\gtrsim$70 Mm) coronal cloud filament.  Here, we examine closer the response of the lower atmosphere to the impact of this downflow. 

\section{Observations and data reduction} \label{sec:obs}

\subsection{DST/FIRS Observations}

Spectropolarimetric observations covering the \ion{He}{1} 10830 \AA{} triplet were acquired by the Facility Infrared Spectropolarimeter \citep[FIRS:][]{jaeggli2010} at the 0.76 m Dunn Solar Telescope on 19 September 2011.  The target was the large sunspot near N23W11 ($\mu$=$0.92$).   While this is the same data set discussed by \citet{schad2016}, we have reprocessed the raw data to better address spectral and spatial scattered light contamination (see Section~\ref{sec:straylight}).

Figure~\ref{fig:overview} shows an overview of the single scanned FIRS data set used here and the downflow event of interest.  Between 16:29 and 17:01 UTC, FIRS conducted a 218 step slit-scan with the slit axis oriented $41\fdg3$ clockwise relative to the solar meridian. One second cumulative integration time per step was distributed among two repeated sequences of four polarimetric modulation states.  The field of view is $64\arcsec \times 75\arcsec$. Sampling along the slit is $0\farcs15$ and the scanning step width is equal to the projected slit width of $0\farcs29$.  Spectral sampling is 38.683 m\mbox{\AA} pixel$^{-1}$ and the spectral range includes -485 to + 600 km sec$^{-1}$ Doppler velocity coverage of the \ion{He}{1} triplet. Numerous spectral lines are observed in this large spectral bandpass (see Table 1 in \citet{schad2016}).  Our analysis concentrates primarily on the \ion{He}{1} signatures, whose rest components lie at 10829.0911 (Tr1), 10830.2501 (Tr2), and 10830.3398 \AA{} (Tr3)---the latter two transitions are blended, in addition to the shallow \ion{Ca}{1} 10838.97 \AA{} line formed in the deep photosphere with a Land\'{e} g$_{\rm eff}$ of 1.5.

Standard reduction methods as per \citet{schad2016} are applied, including detector nonlinearity correction, dark and flat field correction, geometric registration, wavelength calibration, and polarization crosstalk correction.  We forego removal of any residual interference fringes as flat field correction is sufficient over the region of interest; though some high frequency detector noise is removed through a combination of Wiener and Fourier filtering.  As in our previous work, solar image drift during the slit scan is corrected through careful co-registration with a co-operating G-band context imager and linear interpolation of the FIRS data in the spatial domain.  The cumulative image drift is $2\farcs55$ and occurred primarily over the latter 15 minutes of the scan.  It has a minimal impact on the spatial integrity on scales of a few arcseconds as the linearized slit scanning rate is significantly larger than the average image drift rate. 

\subsection{Solar Dynamics Observatory}\label{sec:obs_sdo}

To investigate plasma at temperatures above the ionization temperature of neutral helium, as well as the temporal behavior of the observed downflow, we include analysis of EUV observations from the Atmospheric Imaging Assembly \citep[AIA:][]{lemen2011} instrument onboard NASA’s \textit{Solar Dynamics Observatory} \citep[SDO:][]{pesnell2012}.  We use the 12 sec cadence level 1 data covering 200 minutes of observations between 15:40 UT and 19:00 UT. We also use continuum intensity data (6173 \AA{}) from the Helioseismic and Magnetic Imager \citep[HMI:][]{scherrer2012} to coalign the observations, via cross-correlation techniques, and to address straylight within the FIRS data.  For the latter, we use the PSF corrected HMI data series denoted 'dconS' \citep{norton2017, norton2018}.  Each SDO image is derotated and rescaled to a common plate scale of $0\farcs6$ pixel$^{-1}$ using the aiapy package \citep{barnes2020}.  Time-series data centered on the region of interest are then coaligned by compensating for the mean synodic solar rotation rate using tools available in the sunpy package \citep{sunpy_community2020}. 

\begin{figure}
    \centering
    \includegraphics[width=0.45\textwidth]{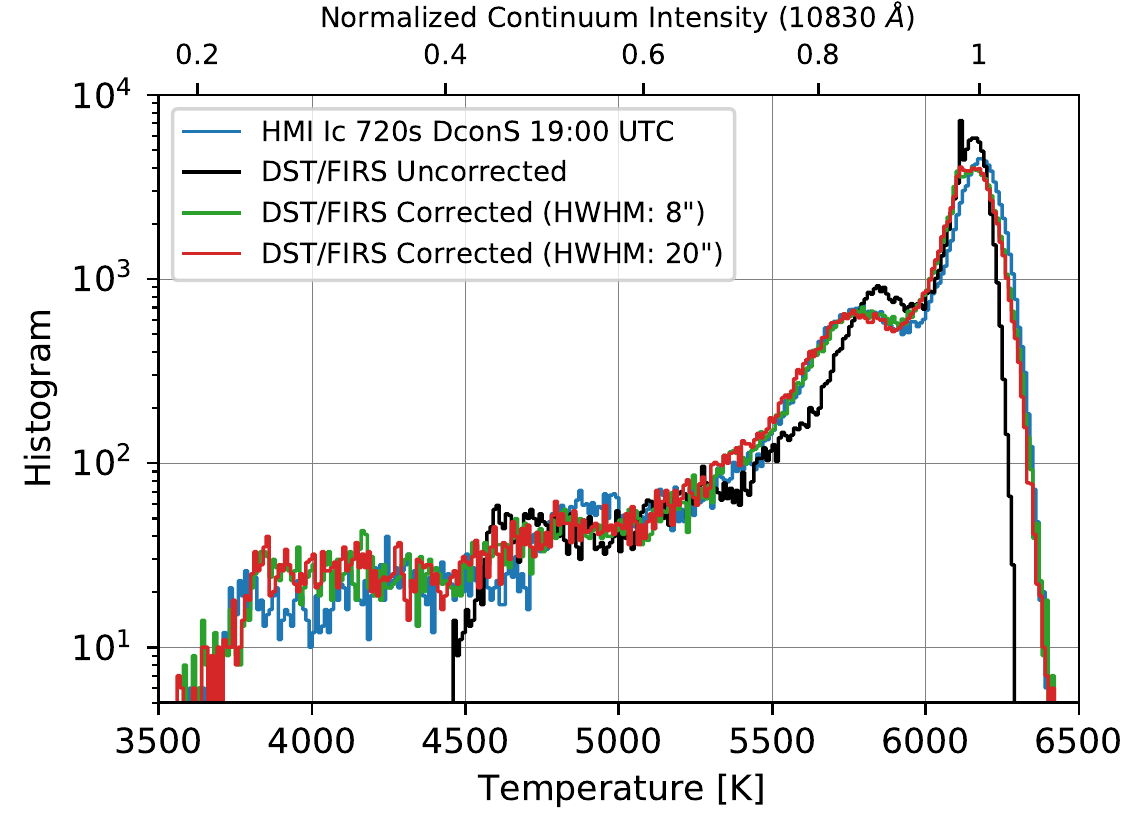} 
    \caption{Histograms of the continuum brightness temperature across the FIRS field-of-view (Fig.~\ref{fig:overview}) for the uncorrected and deconvolved FIRS data as compared to the SDO/HMI PSF-corrected data.} 
    \label{fig:temp_hist}
\end{figure}

\begin{figure}
    \centering
    \includegraphics[width=0.45\textwidth]{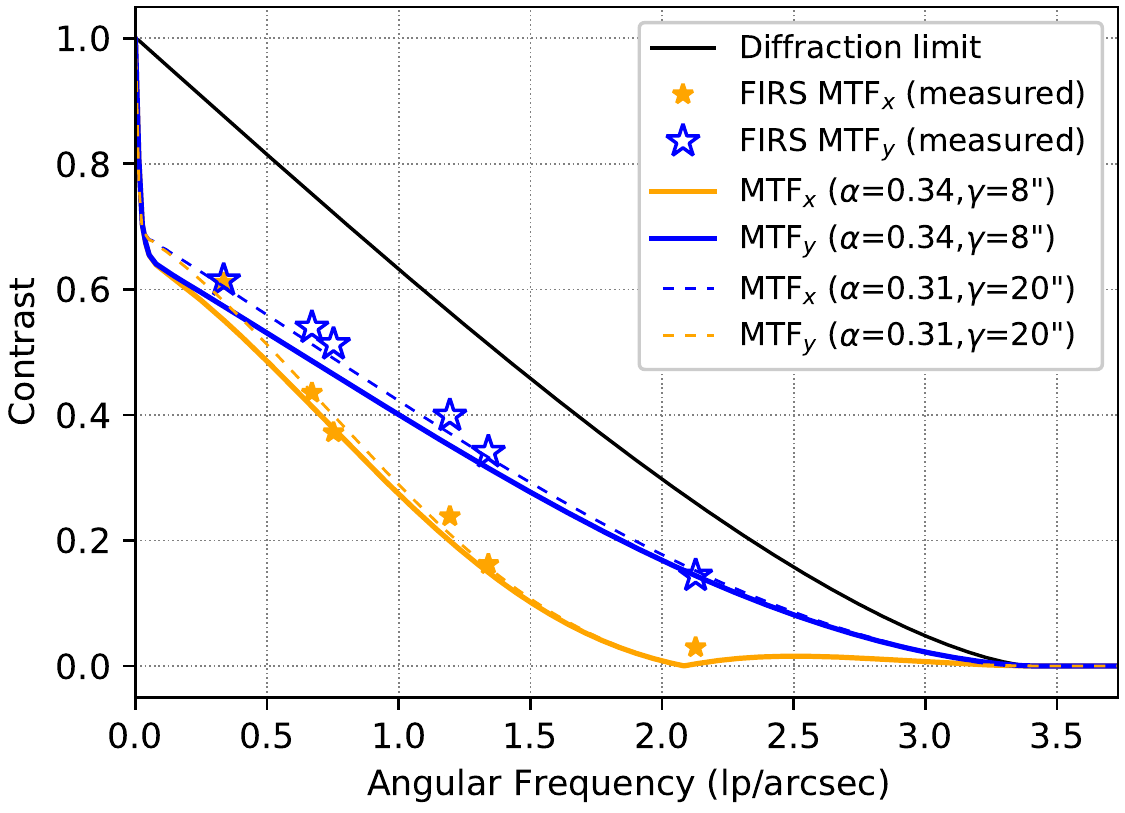} \\
    \includegraphics[width=0.45\textwidth]{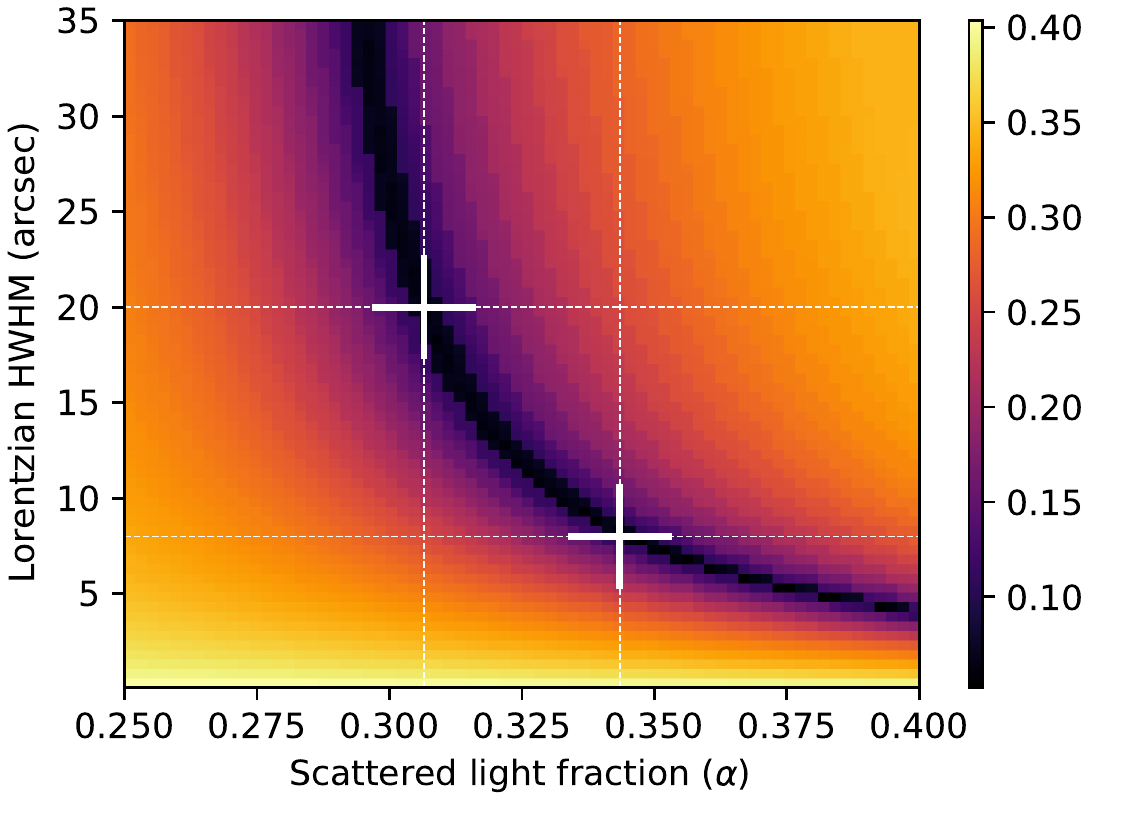} 
    \caption{(top) The FIRS modulation transfer function (\textit{i.e.} contrast measured as a function of line-pairs per arcseconds on sky) values, along the slit (y) and stepping direction (x), inferred from Air Force Resolution Target calibrations at select angular frequencies and compared to the two optimized and selected model PSFs used in this work.  Solid and dashed colored lines give the MTF of the narrow and wide PSF solutions, respectively.  (bottom) The value of the objective function (with arbitrary units) used to optimize the model PSF with the narrow and wide solutions indicated with cross-marks.}
    \label{fig:psf_mtf}
\end{figure}

\subsection{DST/FIRS Straylight Treatment} \label{sec:straylight}

Instrumental straylight negatively impacts the relative spectrophotometry of the FIRS observations, particularly within the sunspot umbra where we concentrate this study. Similar to \citet{borrero2016}, we identify at least two stray light components: (1) a spectrally dispersed ``veil'' component, and (2) a spatially scattered component.  Following the technique of \citeauthor{borrero2016}, we estimate the spectral straylight fraction using flat-field observations to be 7.5\% of the observed spectral intensity while the spectral resolving power is $\approx$100,000.  All FIRS observations are then corrected for this presumed flat spectral stray light component using Equation 4 of \citet{borrero2016}.

Correction of the spatially scattered light is more difficult as we do not have complete knowledge of the point spread function (PSF).  We make relevant approximations using available calibration data optimized through comparisons with straylight-corrected HMI data. Figure~\ref{fig:temp_hist} compares the histograms of the brightness temperature inferred in the continuum across the observed FIRS field-of-view with that of a co-aligned SDO/HMI 'dconS' data product nearest in time, which is a 720 sec temporally smoothed image at 19:00 UT, approximately 2 hours after the FIRS scan.  The local quiet sun brightness temperature at 6173 \AA{} and 10830 \AA{} are both $\approx 6125 K$ \citep{cox2000}, and the sunspot temporal evolution does not play a significant role in the temperature distribution over this time period.  Inferred temperatures in the sunspot umbra are approximately 900 K higher in the uncorrected FIRS data compared to HMI.  

To facilitate correction of the FIRS data, we adopt and optimize an azimuthally symmetric PSF modeled as the sum of a diffraction-limited Airy function (P) and a Lorentzian function (L), further convolved with a 2D discrete instrument sampling profile (D), given by
\begin{equation}
    \rm PSF(x,y) = \rm D(x,y) *[(1-\alpha)P_{Airy}(r; \lambda) + \alpha L(r; \gamma)] \label{eqn:psf}
\end{equation}
where $\alpha$ denotes the scattered light fraction and $\gamma$ is the Lorentzian half width at half maximum (HWHM).   The observations were actively corrected during acquisition by the High-Order Adaptive Optics system \citep{rimmele2004}, which partially justifies the adoption of the diffraction limited Airy function.  

Further constraints on the model PSF are provided by narrow FIRS calibration slit scans of an Air Force resolution target placed in an intermediary focal plane, downstream of the main telescope optical tube. Although this excludes the atmospheric and telescopic contributions to the straylight, this calibration provides measurements of the post-focus performance at moderate angular frequencies (as measured in line pairs per arcsec).  Figure~\ref{fig:psf_mtf} shows the measured contrast values (\textit{i.e.} the modulated transfer function or MTF) in the scanning direction (x: gold stars) and along the slit (y: blue stars).  We first point out that the measured MTF is not symmetric, but instead the contrast is comparatively reduced in the scanning direction, which we model by increasing the effective sample width along the scanning direction by a factor of 1.6 to 0.463\arcsec.  We speculate incorrect timing between the image scanning and data acquisition is responsible.

Within the MTF plot are two different model PSFs described by Equation~\ref{eqn:psf}.  These PSFs are optimized solutions determined using an objective function that seeks a match between the HMI brightness temperature histogram and that recovered by FIRS after spatial deconvolution.  We use 10 iterations of the Lucy–Richardson procedure for deconvolution \citep{richardson1972, lucy1974}. Figure~\ref{fig:psf_mtf} (bottom panel) shows that a ridge of possible solutions exist for the two free variables of our model ($\alpha$,$\gamma$).  The two we select for further comparison are a narrow solution ($\alpha,\gamma = 0.34, 8\arcsec$) and wide solution ($\alpha,\gamma = 0.31, 20\arcsec$), which we use to evaluate the robustness of the results to changes in the modeled PSF.  In the top panel, these are respectively the solid and dashed colored curves.  Solutions with a scattered component narrower than $\gamma=8\arcsec$ are unlikely based on the width of the solar limb profiles observed at other FIRS bandpasses by \citet{jaeggli2012}. Both selected PSFs lead to temperature histograms (Figure~\ref{fig:temp_hist}) that compare well with HMI and have an MTF comparable with the measured values at intermediate frequencies (top panel of Figure~\ref{fig:psf_mtf}). 

Using the two PSF models, we deconvolve the FIRS polarized spectra on an individual wavelength basis and evaluate their differences before detailed analysis below.  This approach provides a more pristine view of the observed phenomena otherwise altered by a large fraction of stray light.  We introduce the observations in the next section and then go on to show the differences in the corrected profiles when using the different PSF models in Section~\ref{sec:spec_profiles}.  We ultimately find them to be of minor consequence, which is why we concentrate our description on the data corrected using the narrow-PSF model.

We also must recognize the limitations of our approach to the PSF correction.  The true PSF is not fully constrained, and we are deconvolving slit-scanned reconstructed images acquired over tens of minutes, which is longer than the characteristic dynamical time scales probed by the \ion{He}{1} spectral lines.  Both may introduce artifacts in the deconvolved signal.  Therefore, we pursue below alternate analysis methods to assess the uncertainty in our results introduced by the stray-light contamination (see Section~\ref{sec:fits_non_deconv}).

\section{He I Data Analysis}

\begin{figure*}
    \includegraphics[width=0.99\textwidth]{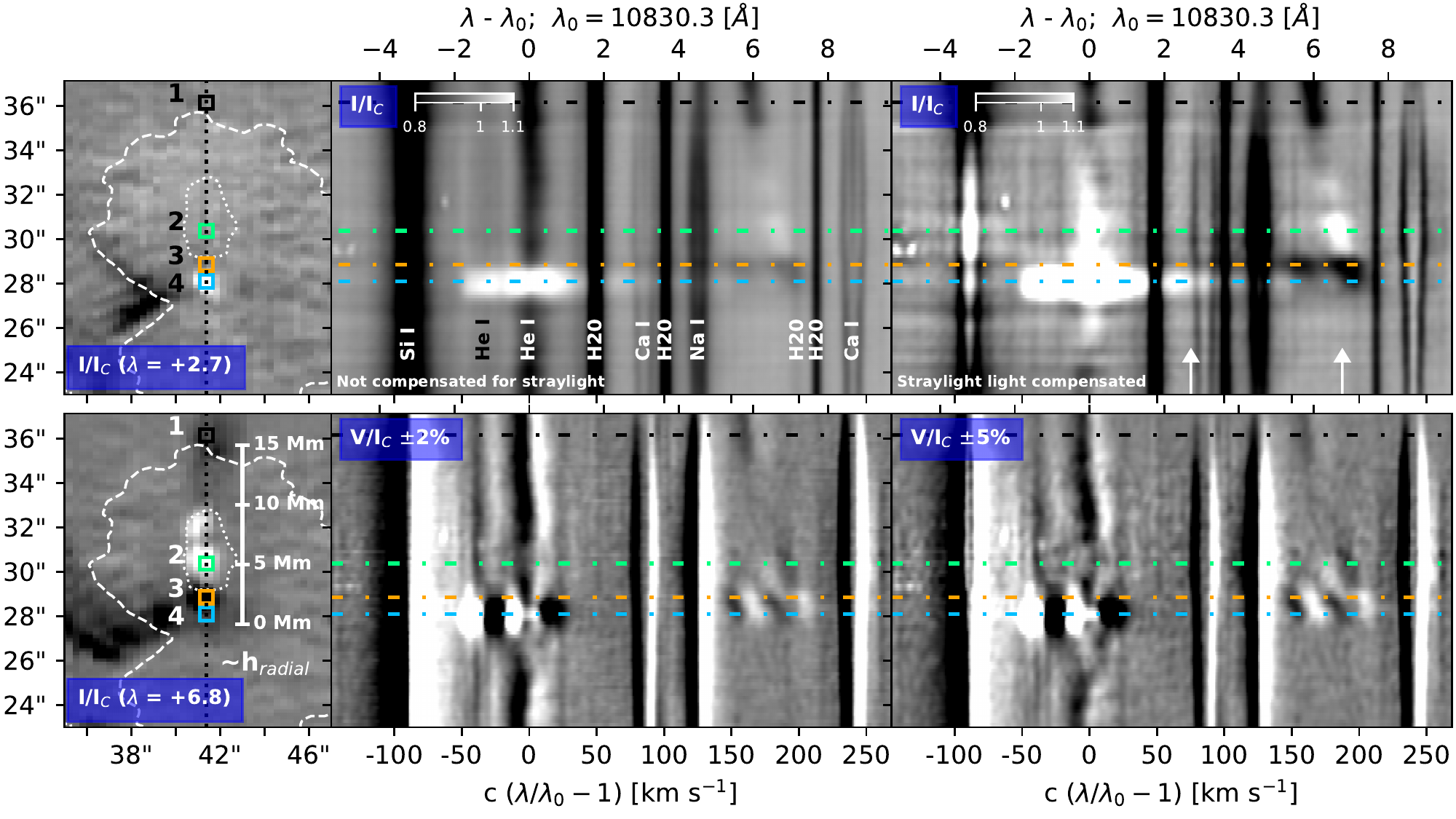}
   \caption{DST/FIRS Stokes I and V image spectra along the footpoint of the draining coronal loop.  The spatial images in the left column are the same as panels (h) and (i) in Fig.~\ref{fig:overview}, and the spectra at right are extracted along the dotted lines.  Both the uncorrected (middle) and PSF-deconvolved (right) spectra are displayed.  Arrows in the upper right panel indicate the wavelength location of the images at left. The color scales linearly for Stokes I, which is normalized to the local continuum for each spectra row, between 0.8 and 1.1 to enhance features; though, it saturates in the absorption lines and bright emission spectra near 10830 \AA{}.  The absorption and emission features near 10836 \AA{} are highly redshifted components of neutral Helium with speeds between 160 and 195 km s$^{-1}$.}
    \label{fig:image_spec}
\end{figure*}

\subsection{Signatures of the supersonic downflow and its impact}

In Figure~\ref{fig:overview}, spatial maps of the deconvolved FIRS intensity data (using the narrow PSF model) are shown at different wavelengths and for a zoomed in region of the sunspot umbra.  The uncorrected data, as presented in \cite{schad2016}, shows qualitatively equivalent features.  The field-of-view includes the area north and east (see cardinal directions in panel(e)) of the dark umbral core where large filament structures are seen in panel (b) near the rest wavelengths of the \ion{He}{1} triplet. Distinct from these filaments and at much larger redshifts (67 km s$^{-1}$), panel (c) shows multiple channels of absorption left of the sunspot in the image directed towards the sunspots, most notably near X,Y $=15\arcsec,65\arcsec$. \citet{schad2016} traced 23 of these channels, each of which exhibit acceleration of downflowing material converging primarily towards the inner boundary of the photospheric penumbra.  

The most prominent downflow channel, which is the subject of the rest of this work, extends from the upper left of the field-of-view, curves downwards, and traverses (in projection) the darkest part of the sunspot umbra, as seen in panels (d) and (i) at redshifts of 179 km s$^{-1}$.  As it traverses the umbral core, the redshifted feature transitions from absorption to emission and then back into absorption. The high speed emission profiles reside within the bounds of a contour line tracing continuum intensities below 0.23 normalized to the quiet Sun. Meanwhile, the downflowing material persists in these monochromatic images up to redshifts of 206 km s$^{-1}$ (panels e and j) wherein only a small absorptive feature approximately $1\arcsec$ in size is observed near X,Y = $42\arcsec,28\arcsec$, within the sunspot umbra and inside of the 0.28 normalized continuum intensity contour.  Coincident with the small patch of highly redshifted absorption in panel (j) is a small kernel of emission, similar in spatial size, and strongest near the rest wavelengths of the \ion{He}{1} triplet, as seen in panel (g).  We refer to this kernel, which is highlighted by overlapping contours in panels (g) and (j), as the terminus of the downflow within the umbra.  We do not find evidence for significantly blue-shifted emission within this kernel nor any obvious signature in the photospheric continuum; however, \ion{He}{1} emission in this kernel is evident at large redshifts, as shown in panel (h) at a redshift of 67 km s$^{-1}$, which corresponds to red line wings in the spectra discussed below. 

\begin{figure*}
    \centering
    \includegraphics[width=0.99\textwidth]{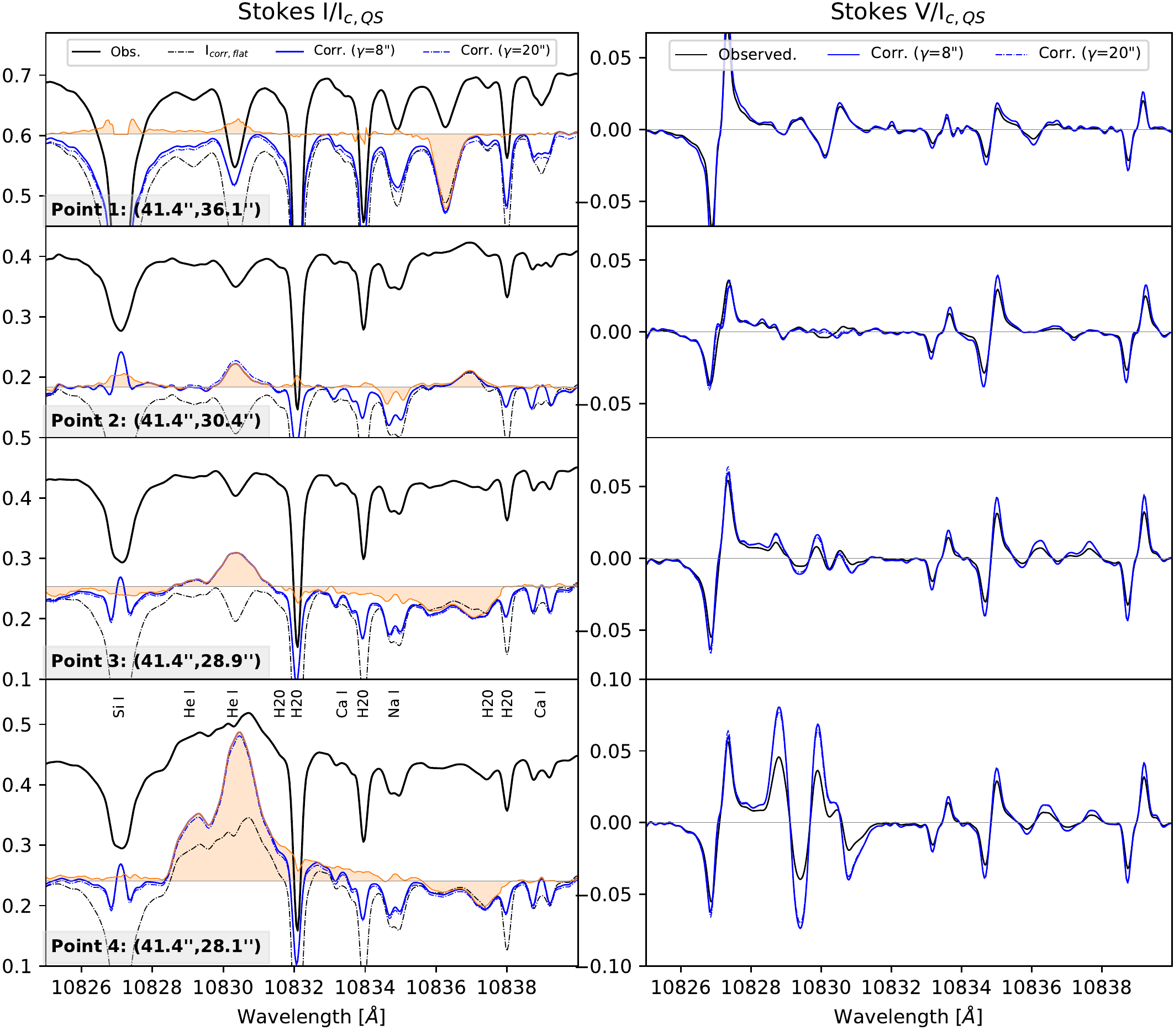} 
    \caption{The original uncorrected data (solid black) and the spatially deconvolved Stokes I and V spectra (blue) extracted from the four selected locations in Fig.~\ref{fig:image_spec}.  Two versions of the corrected data are shown, one for each selected PSF model.  The dash-dotted black lines show the original intensity profile corrected assuming a spectrally-flat straylight component, while the filled orange curves shows the corrected ($\gamma=8\arcsec$) profiles for which an estimate of the background absorption profiles has been removed.  See text for more details.}
    \label{fig:spec_backg}
\end{figure*}

\subsection{Image spectra}

The spectrograph slit was fortuitously well aligned with the lower segment of downflow near $X{\sim}41.1\arcsec$. Figure~\ref{fig:image_spec} displays the corresponding observed and narrow-PSF-corrected Stokes I and V image spectra. The high speed \ion{He}{1} absorption immediately above the outer umbral edge (Y$>35\arcsec$) in the lower left panel corresponds to the satellite component centered near 10836.3 \AA{} (${\sim}166$ km s$^{-1}$).  It is observed in both Stokes I and V, between photospheric \ion{Na}{1} lines near 10834.8 \AA{} and telluric water lines near 10838 \AA{}.  The weaker (Tr1) helium transition is not immediately evident. Proceeding downwards (along the slit) into the umbra, this satellite component of \ion{He}{1} shows the transition described above from absorption to emission and back to absorption, along with the corresponding change in sign of the Stokes V profile.  Its Doppler redshift also continues to increase. 

In the area of highly redshifted absorption (X,Y = $41.1\arcsec,28.7\arcsec$), strong signatures of what is presumed to be the blueward Tr1 \ion{He}{1} transition appear. Immediately below these signatures (X,Y = $41.1\arcsec,27.9\arcsec$) the high speed absorption weakens and strong emission profiles near the \ion{He}{1} rest wavelengths appear with red wings extending to Doppler shifts nearly as high as the proceeding high speed absorption. Note the color scale saturates for intensities 10\% larger than the local continuum to enhance the weaker features.  In Stokes V, strong Zeeman-like profiles exist near the rest wavelengths with an opposite sign relative to the background \ion{He}{1} spectra in the outer umbra. In the extended red wing, though, the emission profiles do not show pronounced Stokes V signals (\textit{i.e.} between 10832 and 10835 \AA{}).

The primary spectral features of this downflowing event are common to both the uncorrected and deconvolved data; however, the straylight correction can significantly alter the line profile shape.  The corrected \ion{Si}{1} 10827.09 \AA{} line within the umbral core shows reversals in the line core, both in Stokes I and V, within the corrected data. \citet{orozco_suarez2017} reported the initial discovery of these reversals in straylight corrected data from the German GREGOR telescope.  In addition, and as expected, the observed unshifted ($\pi$) component of the intensity profile in the strongly Zeeman split photospheric lines is largely reduced upon straylight correction.  Less expected, however, are the \ion{He}{1} emission signals only found in the straylight corrected data near the rest wavelengths ($29\arcsec < y < 33\arcsec$),which also show disrupted complex V profiles.  While umbral \ion{He}{1} profiles have shown complex profiles in previous observations (see, \textit{e.g.}, \citet{lites1986} and \citet{centeno2005}), emission profiles are not typically observed.  It is possible that these emission signals are real; though, we currently cannot discount that they are artifacts introduced by the straylight correction. 

\subsection{Spectral profiles}\label{sec:spec_profiles}

Figure~\ref{fig:spec_backg} displays the observed (solid black) and corrected spectral profiles (solid and dashed blue) sampling key locations along the extent of the channel footprint, as indicated by dash-dotted lines in Figure~\ref{fig:image_spec} and labeled with indices 1 through 4 (top to bottom in each Figure). We select profiles sampling (1) the high speed absorption immediately outside the umbra, (2) the strongest redshifted emission, (3) the largest degree of high speed absorption in the umbra, and (4) the strongest emission at rest wavelengths at the downflow terminus.

To better illustrate the effects of the deconvolution on the spectral line shapes, we compare the deconvolved profiles with the original intensity profiles that are first offset and scaled assuming a spectrally-flat straylight component.  This proceeds as follows.  First we write the observed Stokes vector (S=\{I,Q,U,V\}) as a linear combination of the true vector and a scattered profile:
\begin{equation}
S_{obs} = (1-f)S_{true} + f S_{scat}, \label{eqn:comb}
\end{equation}
where, if the scattered light intensity profile is normalized by its continuum,
\begin{equation}
f = \frac{I_{C,obs} -I_{C,true}}{1 - I_{C,true}}
\end{equation}
with $I_{C,obs}$ and $I_{C,true}$ being the observed and true continuum intensities. Approximating $I_{C,true}$ as that obtained from the PSF-corrected data, $f$ can be directly determined.  Then, in the case of spectrally-flat scattered component ($I_{scat} = 1$), we can use Equation~\ref{eqn:comb} to obtain the offset and scaled profiles as 
\begin{equation}
    I_{corr,flat}(\lambda) = \frac{I_{obs}(\lambda) - f}{1-f}, \label{eqn:I_corr_flat}
\end{equation}
which is plotted in Figure~\ref{fig:spec_backg} for each selected profile using dash-dotted black lines. 

Also in the figure, we show a version of the ($\gamma = 8\arcsec$) PSF corrected intensity profiles for which the contribution of the background absorption signals (\textit{i.e.} those not associated with the downflow) are estimated and removed.  These are the orange curves whose areas relative to the continuum are also filled in orange.  As in Section~\ref{sec:model_fits}, the background signals are estimated by first normalizing each profile along the slit by the local continuum and clipping the maximum value to 1, which removes the emissive contribution.  We then median filter these continuum-normalized profiles along the slit with a kernel width of 25 pixels for wavelengths shorter than 10834\AA{} and 100 pixels otherwise.  The increased kernel width is necessary at higher redshifts as the signal of interest is present along a greater portion of the slit.  The filtered and normalized profiles are then removed from the corrected profiles through division.  In Figure~\ref{fig:spec_backg} this process visually enhances the signatures of the downflow relative to the background spectra.

The corrected line shapes in Figure~\ref{fig:spec_backg} show only minor differences between the two selected model PSFs.  We conclude the straylight correction is not overly sensitive to the slope of the modeled PSF wings.  For all further analysis, we adopt only one of the deconvolved data sets, the narrow PSF version ($\gamma$=$8\arcsec$).  The line shapes of the photospheric lines share similarities with those reported by \citet{orozco_suarez2017}; in particular, the \ion{Si}{1} line core intensity is enhanced by ${\sim}0.05$ relative to the local continuum within the darkest pixel shown (point \#2, $I_{C}/I_{C,QS} \approx 0.19$), and the core of the \ion{Ca}{1} 10839 \AA{} line is raised to the continuum level. 

The profiles of the deconvolved, high-speed components of \ion{He}{1} (between 10836 and 10837.5 \AA{}) are slightly enhanced and/or narrowed relative to those employing a spectrally-flat straylight profile (\textit{i.e.}, Equation~\ref{eqn:I_corr_flat}), which is expected given the deconvolution should mitigate the influence of the significant gradients in the Doppler shift of this component along the slit that are smeared by the PSF.

Near the \ion{He}{1} rest wavelengths, the point \#1 helium profile appears consistently in absorption for the observed and corrected data.  For point \#2, which is displaced ${\sim}2\arcsec$ from the loop footpoint, the corrected data display a weak emission profile, and the original Stokes V signal is weakened, leaving a complex profile likely dominated by noise introduced during the deconvolution.  Emission above the continuum in point \#3, and especially point \#4, is evident in the uncorrected data and becomes pronounced in the corrected data.  In these locations, the Stokes V profiles are also enhanced, and the sign of the blue-ward, anti-symmetric lobe is opposite of the absorption profile in point \#1, consistent with the emission profiles and the expectation that the polarity of the magnetic field remains constant across the region.  

By comparing the observed and corrected profiles for point \#4, especially near the \ion{He}{1} rest wavelengths, it is apparent that the observed profile is the additive combination of a asymmetric emission profile and a scattered absorptive component (see also Section~\ref{sec:fits_non_deconv}).  The peak of the emission is enhanced by approximately 0.2 in the deconvolved data, in units of the local quiet sun continuum, when compared to the profile scaled using Equation~\ref{eqn:I_corr_flat}. This is not unexpected given the median \ion{He}{1} profile across the field-of-view exhibits ${\sim}$30\% absorption. Three maxima within the observed emission are transformed into two, which is more compatible with the characteristics of the \ion{He}{1} 10830 \AA{} triplet.  Furthermore, the anomalous Stokes V reversal near the core of the blended Tr2+Tr3 component (10830.3 \AA{}) is removed. 

\subsection{He I Slab Model Fitting}\label{sec:model_fits}

We now perform spectropolarimetric inversions of the observed \ion{He}{1} spectra to extract a range of physical parameters.  Each selected Stokes spectrum is analyzed with custom multi-line models.  For the \ion{He}{1} triplet, we use forward synthesis modules provided by the {\sc Hazel}-2 code \citep{asensio_ramos2008}, which can treat the full quantum formation of the \ion{He}{1} polarized signals including the Hanle, Zeeman, and Paschen-Back effects.  {\sc Hazel}-2 solves for the radiative transfer in a constant-property slab model using a five term atomic model for triplet state helium taking only radiative transitions into account.  The emergent intensity is determined via an analytical solution to the polarized radiative transfer equation given by 
\begin{equation}
{\bf I}={\rm e}^{-{\mathbf{K}^{*}}\tau}\,{\bf I}_{0}\,+\,\left[{\mathbf{K}^{*}}\right]^{-1}\,
\left( \mathbf{1} - {\rm e}^{-{\mathbf{K}^{*}}\tau} \right)\,\beta \mathbf{S}. \label{eqn:slab_model}
\end{equation}
$\bf I_{0}$ is the Stokes vector incident on the boundary of the constant-property slab, furthest from the observer, and along the same line-of-sight.  $\mathbf{K}^{*}$ is the propagation matrix normalized by its first element ($\eta_{I}$), $\tau$ is the optical thickness of the slab, and $\mathbf{S}$ is the source function vector.  Each parameter, other than $\beta$, is wavelength-dependent.  Values reported below for the slab optical thickness refer to $\Delta\tau_{red}$, the optical thickness at center of the blended red component (Tr2,Tr3).  

$\beta$ is an adhoc scalar parameter introduced to scale the source function to account for deviations from the radiatively controlled five term model atom.  As noted by \citet{asensio_ramos2008}, the five term model does not consistently treat the population mechanism of the helium triplet, which in typical chromospheric conditions is thought to be photoionization of singlet helium by coronal EUV radiation with subsequent recombination into the triplet state \citep{avrett1994}.  Recombination into the triplet state can also occur subsequent to collisional excitation/ionization of singlet helium or through the cooling of ionized coronal plasma.  However, the ionization and recombination processes are slow in comparison to the bound-bound radiative transitions within the triplet, and bound-bound transitions via electron collisions are typically negligible at chromospheric densities.  Here, though, they likely do play a role in certain cases, for which we adopt $\beta$ as a free model parameter.  In the optically thin case, however, $\beta$ is ill-constrained and we cannot include it as a free parameter.  This can be shown, following \citet{trujillo_bueno2007}, by using an approximated form of Equation~\ref{eqn:slab_model} and neglecting anomalous dispersion, wherein the optically thin limit yields for Stokes I: 
\begin{equation}
    I(\tau) \approx I_{0} + \tau (\beta S_{I} - I_{0}), 
    \label{eqn:opt_thin}
\end{equation}
and thus $\tau$ and $\beta$ are compensating variables.  

\subsubsection{Height estimates}

The source function $\mathbf{S}$, in the collision-free case, is governed by the incident radiation field, which is a strong function of location and height relative to the sunspot.  Thankfully, stereoscopic imaging analysis in \citet{schad2016} provides reliable constraints on the geometry of the observed flow, which subsequently allows calculation of the incident radiation field.  The flow's lower portion ($h{\lesssim}20$ Mm) is directed primarily radially and is thus inclined by ($\arccos(\mu) \approx$) $23^{\circ}$ relative to the line-of-sight.  As a relative height reference, Figure~\ref{fig:image_spec} provides an inset axis that accounts for the projection of the radially oriented downflow, and assumes a zero height at the loop terminus (point \#4).  We derive estimated heights for use in our models via this scale. However, the absolute height of the loop terminus cannot be precisely inferred from the FIRS observations; although, it is reasonable to assume it is fairly low in the atmosphere.  \citet{schad2016} triangulated heights of $\approx 3 \pm 3.5$ Mm at the lowest point of the flow; however, the error is likely larger than reported once co-alignment and pointing errors are further taken into account.  Our presumption that the height of the helium signals at the loop terminus is in the low atmosphere is partially justified \textit{a posteriori} based on the inferred field strengths and spectral formation details discussed below, though we also must discuss the influence of these uncertainties on the results. 

\subsubsection{The incident radiation field}\label{sec:rad_tensor}

\begin{figure}
    \centering
    \includegraphics[width=0.45\textwidth]{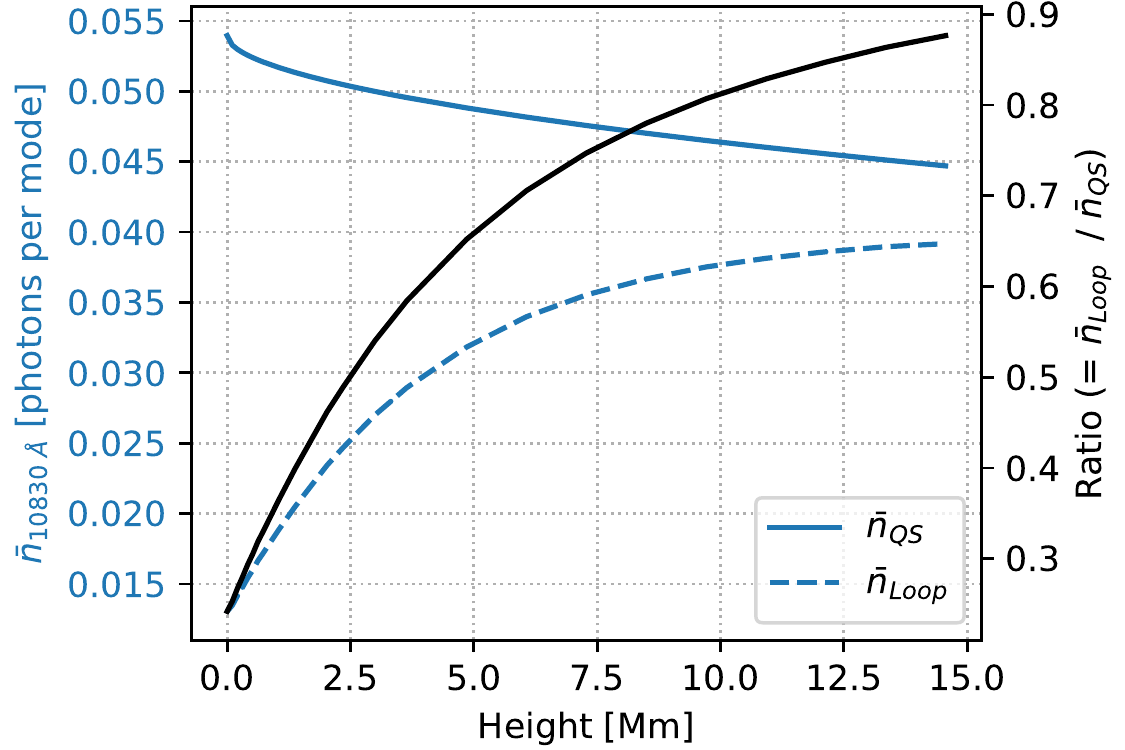}
    \caption{The mean radiation field intensity of the 10830 \mbox{\AA} continuum as a function of height above the photosphere for the quiet sun ($\bar{n}_{QS}$) and radially above point \#4 ($\bar{n}_{Loop}$) in Fig.~\ref{fig:spec_backg}.}
    \label{fig:rad_tensor}
\end{figure}

Neglecting any variations in the continuum formation height, we numerically integrate the deconvolved FIRS continuum data (near 10830 \AA{}) to determine the mean intensity along the downflow, quantified as in \citet{asensio_ramos2008} by the number of photons per mode ($\bar{n}$): 
\begin{equation}
    \bar{n}(\nu) = \frac{c^{2}}{2h\nu^{3}} J_{0}^{0}(\nu) =\frac{c^{2}}{2h\nu^{3}} \oint \frac{d \Omega}{4\pi} I(\nu,\mu)
\end{equation}
where $J_{0}^{0}$ is the first irreducible spherical tensor component of the radiation field. The spectral irradiance angular dependence, $I(\mu)$, is sourced from Allen's Astrophysical quantities \citep{cox2000}, and is scaled according to the fractional contrast of the measured continuum intensity.  For locations above the quiet sun, $\bar{n}_{QS}$ decreases slowly with increasing height (Figure~\ref{fig:rad_tensor}). For a column radially above point \#4, however, $\bar{n}_{Loop}$ increases sharply with height as the radiation cone angle encompasses progressively more of the penumbra and quiet sun.  At a height of 15 Mm, $\bar{n}_{Loop}$ is $\approx$ 90\% $\bar{n}_{QS}$. While $\bar{n}_{Loop}$ relative to the quiet sun will vary for each transition wavelength of the 5 term model, we find this effect negligible in our models compared to the overall scale reduction calculated for the 10830 \AA{} transitions, which below is applied to all transitions uniformly. 

Higher order multipoles of the incident radiation field influence the degree of scattering induced polarization, the signatures of which can be found along the coronal downflow at higher heights \citep[see][]{schad2016}.  Here, we ignore the effects of atomic polarization and do not fit the linear polarized spectra.  The observational noise in Q and U within the dark umbra typically dominate what signals are apparent (see Appendix~\ref{appendix:linpol}).  Furthermore, the strength of the Zeeman Stokes V signals, and the estimated low inclination angle of the flow relative to the line-of-sight, suggest adequate magnetic field inferences can be estimated for \ion{He}{1} by fitting only the circularly polarized spectra. 


\begin{figure}
    \centering
    \includegraphics[width=0.45\textwidth]{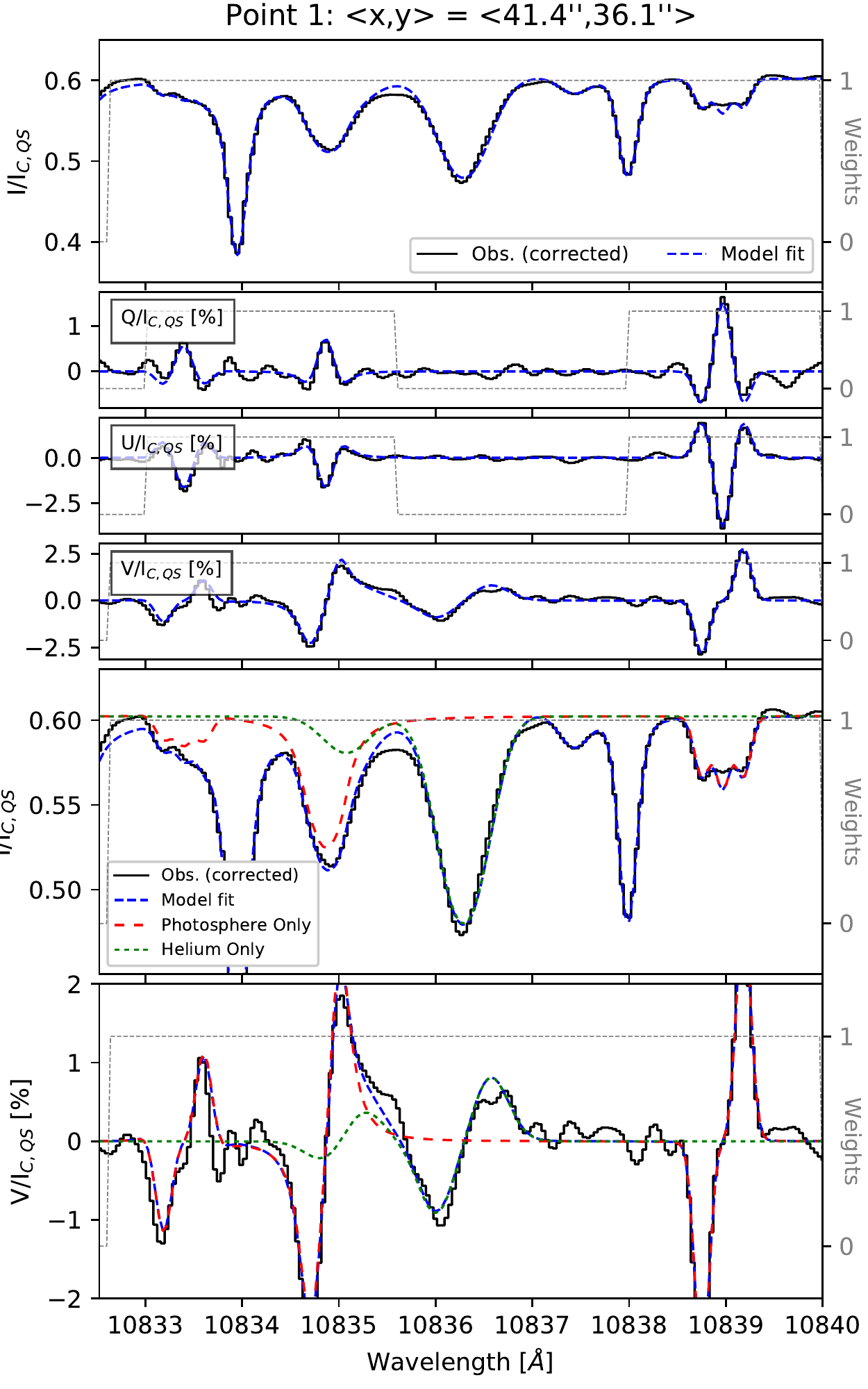}   
    \caption{Inverted slab model fits to the observed high speed \ion{He}{1} absorption at point \#1 near 10836.3 \AA{}. Photospheric and telluric lines are also modeled as discussed in the text.  The bottom two panels show a zoomed in version of I and V with different components of the fit shown separately.}
    \label{fig:inv_1}
\end{figure}

\begin{figure}
    \centering
    \includegraphics[width=0.45\textwidth]{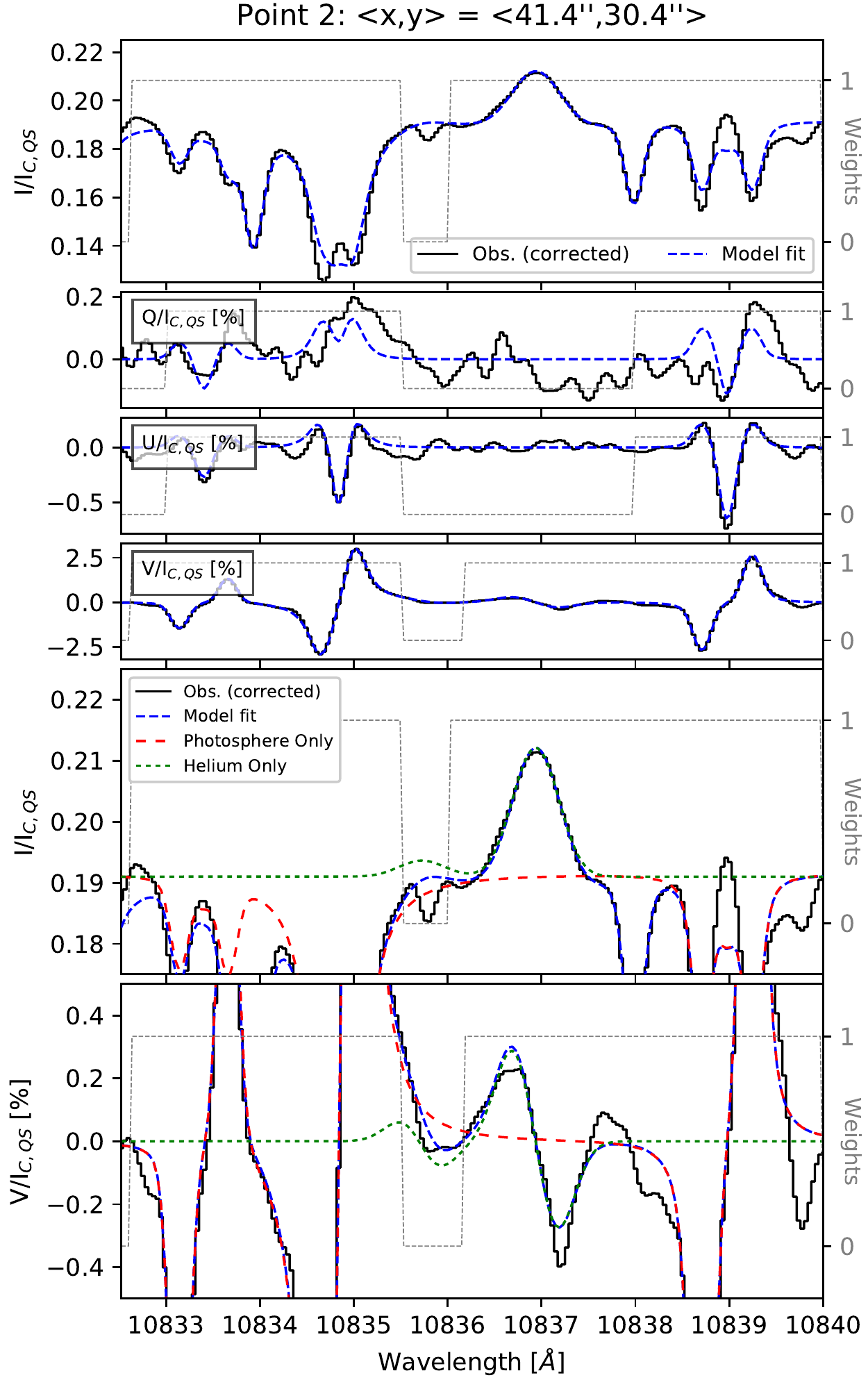}   
    \caption{Same as Fig.~\ref{fig:inv_1} for point \#2 whose helium signal is in emission near 10836.9 \AA{}.}
    \label{fig:inv_2}
\end{figure}

\begin{figure}
    \centering
    \includegraphics[width=0.45\textwidth]{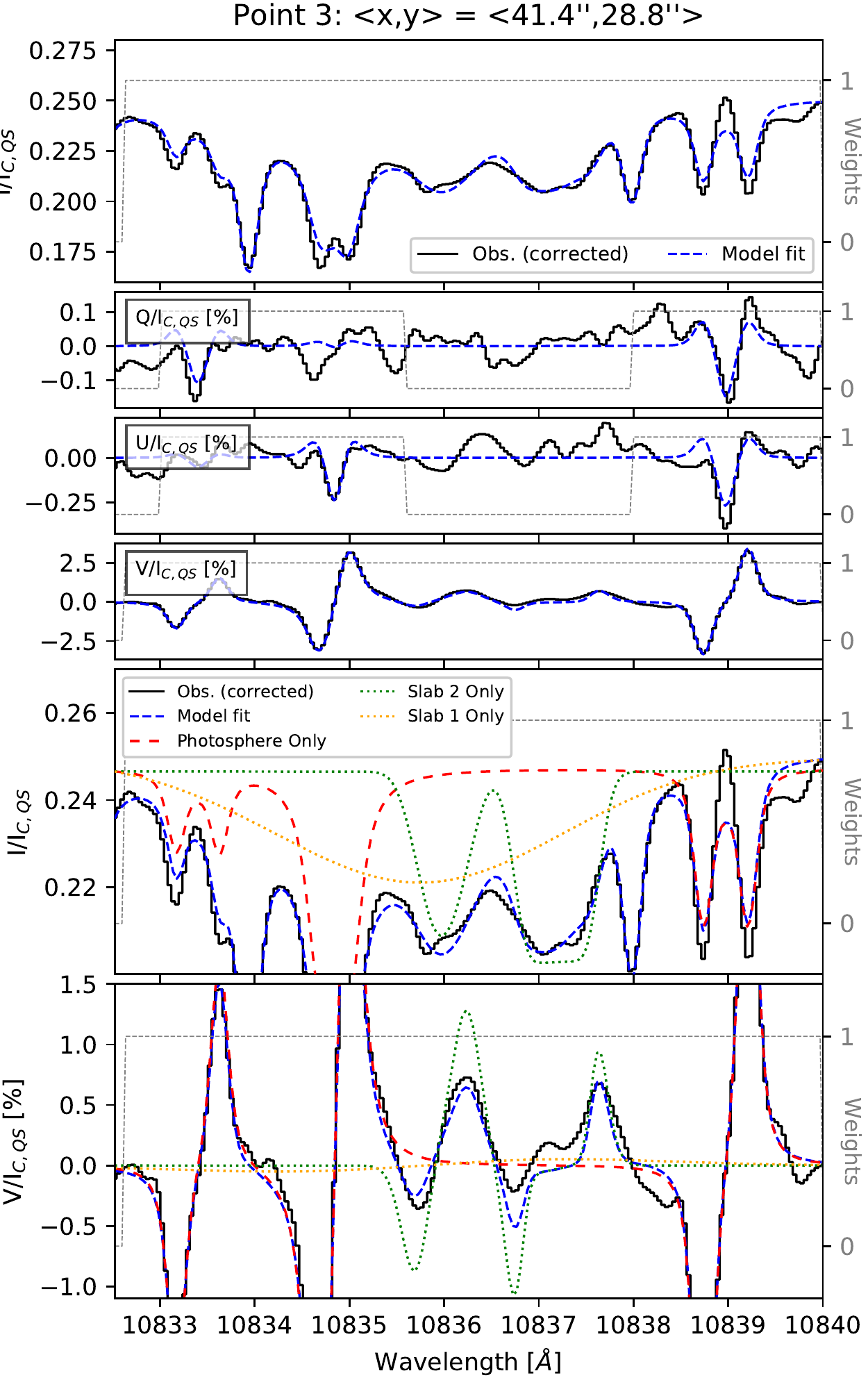}   
    \caption{Same as Fig.~\ref{fig:inv_1} for point \#3.}
    \label{fig:inv_3}
\end{figure}

\subsubsection{Point \#1: Single Helium Slab Model in Absorption}

The absorptive high speed satellite observed at point \#1, centered near 10836.3 \AA{}, can be modeled using a single \ion{He}{1} slab model. The PSF-corrected Stokes spectra and optimized fits are shown in Figure~\ref{fig:inv_1}.  In this case, while the blended Tr2,Tr3 component is resolved, the weak Tr1 \ion{He}{1} transition of the redshifted satellite is only visually apparent in the Stokes V profile and is otherwise dominated by the blend of three \ion{Na}{1} Zeeman-sensitive photospheric lines near 10834.85 \AA{}.  We develop a uniform approach to modeling the background photospheric and telluric lines for points 1, 2, and 3. The five strongest \ion{Na}{1} and \ion{Ca}{1} photospheric lines are modeled using the Unno-Rachkovsky solutions for a Milne Eddington atmosphere \citep{landidegl2004}.  This model then enters Equation~\ref{eqn:slab_model} as  $\bf{I}_{0}$ scaled according to the measured contrast of the background continuum ($I_{C}/I_{C,QS} = 0.6$ for point \#1) and multiplied by the canonical value of the solar spectral irradiance at 10830 \AA{}.  Finally, we use normalized voigt profiles to model telluric line absorption at 10834, 10837.5, and 10838 \AA{}.  

The reduced chi-square statistic, given as the weighted sum of squared deviations per degree of freedom ($\chi^2_{red} = \chi^2 / n_{dof}$), is used as the objective function.  Measurement variances, relative to the quiet Sun intensity, are estimated from the data as $\sigma^{2}(I,Q,U,V) = [0.0036, 0.0009,0.0009,0.001]^{2}$.  Binary fitting weights control which portion of the observed spectra contributes to the fit during the inversion and are shown in each figure. 

For point \#1, there are 32 free variables in the model fit, but ultimately only 4 variables \textit{directly} contribute to the \ion{He}{1} slab model.  These are the longitudinal magnetic field strength $B_{z}$, the slab optical thickness $\Delta\tau_{red}$, the Doppler shift $v_{LOS}$, and the thermal velocity $w_{T}$ governing the line width.  The line profile damping and $\beta$ are held fixed at 10$^{-4}$ and 1, respectively.  The ratio of $\bar{n}$ relative to the quiet sun is fixed at 0.89, as determined in Section~\ref{sec:rad_tensor}, using an estimated height of 15.55 Mm. 

Our inversion strategy includes pre-optimizing the fits of the photospheric and telluric lines using bounded differential evolution optimization, especially as the helium slab model is computationally more costly.  Once initial guesses are adequately determined from this step and manually for the \ion{He}{1} lines, a trust region reflective least squares optimization is used to converge to the best fit solution.  Both optimization tools are part of the SciPy package \citep{scipy_package}.

We achieve satisfactory fits ($\chi^2_{red}= 2.96$) to the point \#1 spectra using the above described model, and the optimized parameters are reported for it and all subsequent fits in Table~\ref{tab:fit_values_master}.  In addition to the final fit, Figure~\ref{fig:inv_1} plots separately the photospheric portion of the model and the helium slab signal under the approximation that the incidence photospheric illumination is line-free.   

\subsubsection{Point \#2: Single Helium Slab Model in Emission} 

Fitting the satellite emission signal near 10837 \AA{} for point \#2 proceeds in the same manner as for point \#1, with the results shown in Figure~\ref{fig:inv_2}. $I_{C}/I_{C,QS}$ is 0.18 while the $\bar{n}$ ratio is fixed to 0.66 (h${\approx}$4.8 Mm). The contribution of a weak unidentified absorption feature at 10835.8 \mbox{\AA}, which is evident in the atlas umbral spectra from \citet{wallace1999}, is weighted out of the fit. A satisfactory fit ($\chi^2_{red}=0.91$) is achieved to the emission without any scaling of the source function ($\beta=1$).

\subsubsection{Point \#3: Two Stacked Helium slabs} 

At point \#3, the redshifted helium signatures are more complex (see Figure~\ref{fig:inv_3}), showing two strongly absorbing features at 10836 and 10837.1 \AA{} in addition to broad absorption near the \ion{Na}{1} lines at 10834.8 \AA{} (see also Figure~\ref{fig:spec_backg}).  The former have characteristics indicative of optically thick material.  In particular, as the opacity increases, the ratio between the Tr1 and blended Tr2, Tr3 intensities will approach unity; meanwhile, Stokes V of only the blended component will develop a flat segment in its core, like that observed near 10837.2 \AA{}.  We, however, found no adequate single slab model and instead adopted a stacked two slab model.  In a stacked slab model, the intensity emergent from one (more-distant) slab is fed into a subsequent slab as $\bf{I}_{0}$ in Equation~\ref{eqn:slab_model}.  We order the slabs by the redshift of the component such that the most redshifted material is described by the slab closest to the observer. That said, for point \#3, the solution is not significantly influenced by the slab order.  In total, 9 model parameters describe the helium profiles: a single common value for $B_{z}$ plus $\Delta\tau_{red}$, $v_{LOS}$, $w_{T}$ and $\beta$ for each slab.  We find that $\beta$ needs to be a free parameter for both slabs to achieve an adequate fit ($\chi^2_{red} = 1.09$). In fact, to account for the broad absorption near 10834.8 \AA{}, a value of $\beta<1$ is required.  Otherwise, we find the opacity will saturate before reaching the degree of observed absorption.  Meanwhile, we limit the fit to one value for $B_{z}$ because the fit is insensitive to the magnetic field of the broader component.  $I_{C}/I_{C,QS}$ = 0.25 and the $\bar{n}$ ratio is fixed to 0.45 (h${\approx}$1.98 Mm). 


\begin{figure*}
    \centering
    \includegraphics[width=0.975\textwidth]{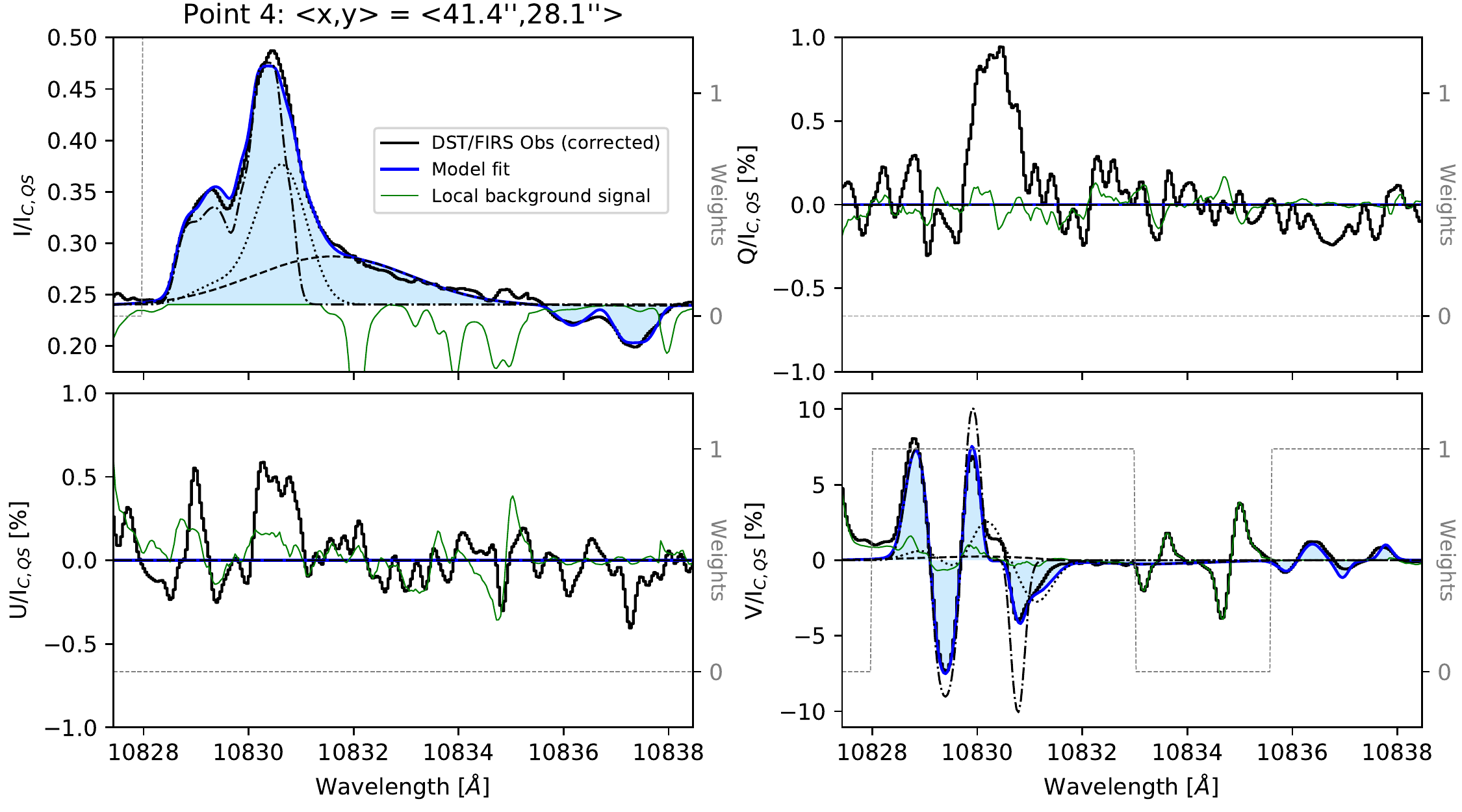}
    \caption{Four component stacked slab model fits to Stokes I and V observed at the downflow terminus at point \#4 in Fig.~\ref{fig:spec_backg} along with the observed Q and U profiles that are not included in the model fit. Thin black dash-dotted, dotted, and dashed lines show the three individual components of the emission signal for the best-fit solution under the assumption that they are independent of the other slabs, which only approximates what is truly modeled in Eq.~\ref{eqn:slab_model}. Thin solid green lines represent the median background signal within the neighboring pixels along the slit.  For Stokes I this background signal has been used to remove the contribution of the non-helium lines prior to model fitting.}
    \label{fig:emiss_fits}
    \label{fig:inv_4}
\end{figure*}

\subsubsection{Point \#4: Fits to the red wing asymmetry} 

For point \#4, our model seeks to reproduce the strong red-asymmetric emission and the redshifted absorption.  The spatial extent of the emission along the slit is only ${\sim}$1\arcsec, and the redshifted absorption signal is separated from the nearby photospheric lines.  This allows us to remove the background lines in Stokes I by dividing the observed profile by the median of the neighboring profiles, which are first normalized and clipped at 1; see Figure~\ref{fig:inv_4}. By removing the background intensity signals and applying spectrally-dependent weights to Stokes V, we can model the observed \ion{He}{1} profiles separate from the photospheric and telluric lines.  

Four stacked helium slabs are necessary to reproduce the observed profile: three slabs for the red asymmetric emission and one for the redshifted absorption component.  The slab ordering relative to the observer is again according to increasing redshift.  The 16 free parameters include $\Delta\tau_{red}$, $w_{T}$, and $v_{LOS}$ for each slab, one value of each $B_{z}$ and $\beta$ that are the same for the three emissive slabs, and also one value of each $B_{z}$ and $\beta$ for the absorptive slab.  In Table~\ref{tab:fit_values_master}, the indices for the slabs are $(1)$ for the absorption slab and $(3,4,5)$ for the emission.  Index 2 is reserved to distinguish the blueward component in point \#3 from the point \#4 emission. 

The assumed height (h${\approx}$0.57Mm) is equal for all components of the signal, as we do not have other constraints.  $I_{C}/I_{C,QS}$ = 0.24 and the $\bar{n}$ ratio is fixed to 0.32.  The quality of the fit ($\chi^2_{red} = 11.78$) is not as good here as for the other selected points, though, the signal is also more complex.  The use of three slabs to model the emission signal is only an approximation of the real atmosphere that likely has strongly varying properties along the line-of-sight.  Improved fits can be achieved with more slabs; though, the number of possible degenerate solutions also amplifies.  In all cases, we have tried to restrict the number of free parameters to those that can be reliably fit. 

The inverted model in Figure~\ref{fig:inv_4} reproduces the main characteristics of the profile.  Recall that Stokes Q and U are not included in the fit.  The Stokes V signal provides important constraints on the widths of the three components needed to reproduce the emission signal and its red wing.  The location of the second slab (index 4 in the table), whose Tr2,Tr3 component peaks near 10830.4 \AA{}, is essential for reducing the V amplitude in the blended component of the slab ahead of it and for creating the V feature near 10831 \AA{}. The only feature the model does not reproduce well is the small offset of the flat portion of Stokes V near 10830.4 \AA{}, which we suspect is a remnant from the imperfect scattered light correction. 

\subsubsection{Point \#4: Fits using alternative straylight corrections}\label{sec:fits_non_deconv}

\begin{figure}
    \centering
    \includegraphics[width=0.45\textwidth]{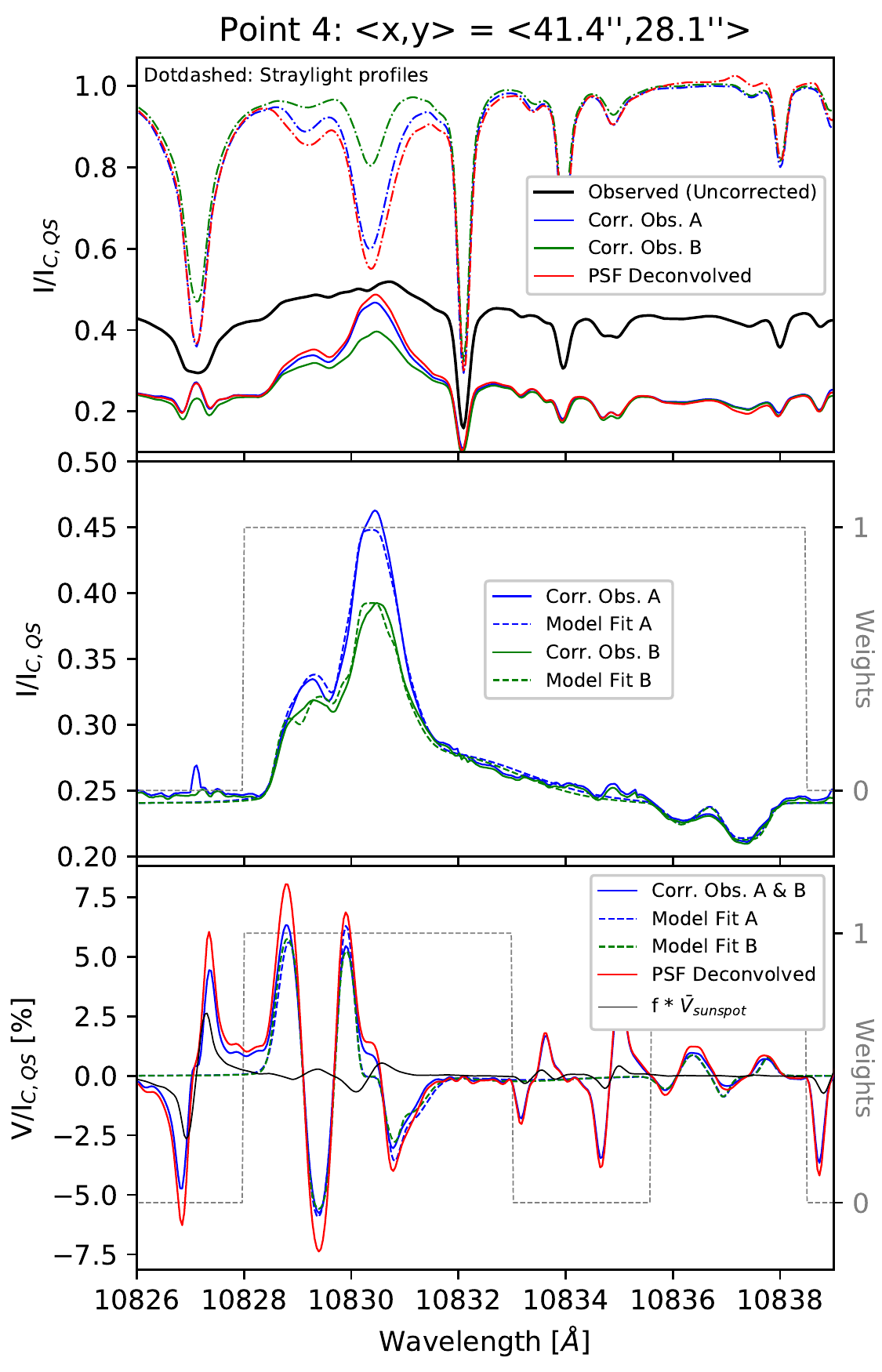}
    \caption{Four component stacked slab model fits of the original point \#4 observations that are corrected for straylight using an alternative method as discussed in the text. (top) The observed profile together with the PSF-deconvolved profile and the two alternative corrected profiles (A and B). Model fits to Stokes I (with the background lines removed) and V are respectively shown in the middle and bottom panels.}
    \label{fig:emiss_fits_wstraylight}
\end{figure}

Finally, although we achieve a satisfactory fit to the deconvolved point \#4 spectrum, given the uncertainties in the PSF correction, we further assess its influence on the derived parameters of the four-slab model.  Using Equation~\ref{eqn:comb}, we first calculate the normalized scattered-light intensity profile which has been removed during deconvolution, shown as the red dash-dotted line in the top panel of Figure~\ref{fig:emiss_fits_wstraylight}. In comparison, the blue and green dash-dotted lines are the first quartile (Q1) and 90th percentile (P90) intensity values by wavelength for all normalized profiles observed across the full FIRS field-of-view.  These represent cautiously wide bounds for the maximum and minimum intensity signal that could be scattered into the observed umbral signal.  The shape and width of each are comparable to the profile removed during deconvolution; although, the absorption depth varies substantially.  Also, note that the PSF deconvolution will not only remove straylight but also locally sharpen the signal, which is why its scattered absorptive component (red dash-dotted line) appears deeper in the figure.  Using Equation~\ref{eqn:comb}, we derive two alternatively corrected profiles, labeled A and B, that make use of the Q1 and P90 profiles as the scattered intensity profile.  For Stokes V, the scattered profile is taken as the mean normalized V profile within the sunspot region (shown in figure multiplied by f), which is small in comparison to the signal in the emission profile. We then carry out the same analysis on these two alternatively corrected Stokes profiles as in the prior section.  The fitted profiles are shown in Figure~\ref{fig:emiss_fits_wstraylight} and their parameters given in Table~\ref{tab:fit_values_master}. One particular difference is that the inferred magnetic field amplitude decreases when the scattered light profile is assumed to have a great amount of helium absorption.  This is expected since the relative degree of circular polarization decreases in that case.  Related changes occur for $\Delta\tau_{red}$ and $\beta$ for the most-distant slab (index 5 in Table) while $v_{LOS}$ and $w_{T}$ are not greatly altered, nor the parameters that describe the red line wing or the fast downflowing component. 

\subsubsection{Summary of Helium Model Fits and Error Estimates}

Table~\ref{tab:fit_values_master} provides the fitted parameters for each spectral fit that has been discussed.  As noted in the caption, the superscript indices for each parameter denote the slab index, which increases with distance from the observer along the line-of-sight, and is further ordered by the Doppler shift of the modeled component.  Slab index 1 is always the most redshifted component of the downflow.  Parameters that are coupled across slabs or fixed are as indicated.  95\% confidence intervals are explicitly calculated for each parameter using the extra-sum-of-squares F-test (see, \textit{e.g.}, \citet{motulsky2004}).  In short, this method determines the significance level of a nested model under the null hypothesis.  Here the nested model is one that reduces the degrees of freedom by one by fixing one parameter to a specific value, which is iteratively changed in both directions from its best-fit value. The significance level is determined for each change using the F-test where the relevant statistic is given by the relative difference in the weighted sum of squared deviations (chi-square) for the best-fit ($\chi^2_{best-fit}$) and the nested test model ($\chi^2_{test}$) divided by the relative difference of degrees of freedom.  Here, this reduces to
\begin{equation}
    F(1,n-m) = \left (\frac{\chi^2_{test}}{\chi^2_{best-fit}} - 1 \right ) (n-m).
\end{equation}
$n$ is the number of observations, and $m$ is the number of free parameters for the best fit case, which is one greater than the model under test.  $F(1,n-m)$ is the F-distribution with numerator $1$ and denominator $(n-m)$.
The significance level is calculated using the cumulative distribution of the F-distribution.  This method provides an estimate for the sensitivity of the fits to changes in a given parameter; though, it does not provide information regarding inter-parameter correlations.  Our models minimize the number of parameters for the helium slabs to reduce the potential for strongly correlated variables.  In the case of point \#4, we expect the uncertainties introduced by the straylight correction to be greater than possible degenerate models. 

\begin{deluxetable*}{lcccc|cc}
\tablecaption{Retrieved parameters using stacked {\sc Hazel}-2 slab models fit to the observed \ion{He}{1} triplet spectra.  The top section contains static aspects of the model and the best fit reduced chi-square statistic.  Superscript indices for each model parameter below denote the slab index, which increases with distance from the observer along the line-of-sight and is furthered ordered by the Doppler shift of the modeled component.  See text for more details. \label{tab:fit_values_master}}
\tablehead{
  \colhead{} & \multicolumn{4}{c}{Inversions of Deconvolved Data}  & \multicolumn{2}{c}{Alternate straylight correction}  \\ 
  \hline
  \colhead{Parameter} & \colhead{Point \#1} & \colhead{Point \#2} & \colhead{Point \#3} & \colhead{Point \#4} & \colhead{Point \#4A} & \colhead{Point \#4B} 
}
\startdata
Estimated Height [Mm]\tablenotemark{a} &   15.55  & 4.81  & 1.98  & 0.57  & 0.57  & 0.57  \\ 
I$_{C}$/I$_{C,QS}$ (obs)   &   0.7  & 0.41  & 0.45  & 0.44  & 0.44  & 0.44  \\ 
I$_{C}$/I$_{C,QS}$ (corr)  &   0.6  & 0.18  & 0.25  & 0.24  & 0.24  & 0.24  \\ 
$\bar{n}$ ratio            &   0.89  & 0.66  & 0.47  & 0.32  & 0.32  & 0.32  \\ 
Num. of slabs              &   1  & 1  & 2  & 4  & 4  & 4  \\ 
$\chi^{2}_{red}$           &   2.96  & 0.91  & 1.19  & 11.78  & 8.39  & 6.55  \\ 
\hline
B$_{z}^{1}$ [G]              & -461$^{+59}_{-60}$    & -841$^{+180}_{-227}$    & -856$^{+113}_{-121}$    & -1351$^{+934}_{-824}$    & -1347$^{+262}_{-274}$    & -1305$^{+238}_{-251}$      \\ 
$\Delta\tau^{1}_{red}$       & 0.68$^{+0.03}_{-0.03}$    & 0.35$^{+0.05}_{-0.05}$    & 11.88$^{+3.64}_{-3.08}$    & 6.05$^{+4.98}_{-3.23}$    & 6.11$^{+1.94}_{-1.75}$    & 6.32$^{+1.83}_{-1.63}$      \\ 
v$_{LOS}^{1}$ [km s$^{-1}$]  & 165.6$^{+0.3}_{-0.3}$    & 183.6$^{+0.7}_{-0.7}$    & 190.6$^{+0.4}_{-0.4}$    & 195.5$^{+1.5}_{-1.9}$    & 195.2$^{+0.6}_{-0.6}$    & 195.2$^{+0.6}_{-0.6}$      \\ 
w$_{T}^{1}$ [km s$^{-1}$]    & 9.06$^{+0.44}_{-0.43}$    & 8.63$^{+0.94}_{-0.91}$    & 7.27$^{+0.5}_{-0.48}$    & 7.52$^{+2.15}_{-1.47}$    & 7.4$^{+0.81}_{-0.74}$    & 7.65$^{+0.7}_{-0.6}$      \\ 
$\beta^{1}$                  & 1 (fixed)         & 1 (fixed)         & 1.09$^{+0.01}_{-0.01}$    & 1.59$^{+0.06}_{-0.08}$    & 1.67$^{+0.03}_{-0.03}$    & 1.66$^{+0.02}_{-0.02}$      \\ 
\hline
B$_{z}^{2}$ [G]              & -                 & -                 & [$=B_{z}^{1}$]    & -                 & -                 & -                   \\ 
$\Delta\tau^{2}_{red}$       & -                 & -                 & 0.13$^{+0.11}_{-0.13}$    & -                 & -                 & -                   \\ 
v$_{LOS}^{2}$ [km s$^{-1}$]  & -                 & -                 & 153.5$^{+4.3}_{-3.8}$    & -                 & -                 & -                   \\ 
w$_{T}^{2}$ [km s$^{-1}$]    & -                 & -                 & 56.1$^{+7.52}_{-6.61}$    & -                 & -                 & -                   \\ 
$\beta^{2}$                  & -                 & -                 & 0.1$^{+0.48}_{-0.1}$    & -                 & -                 & -                   \\ 
\hline
B$_{z}^{3}$ [G]              & -                 & -                 & -                 & [$=B_{z}^{5}$]    & [$=B_{z}^{5}$]    & [$=B_{z}^{5}$]      \\ 
$\Delta\tau^{3}_{red}$       & -                 & -                 & -                 & 0.23$^{+0.03}_{-0.07}$    & 0.2$^{+0.02}_{-0.02}$    & 0.29$^{+0.03}_{-0.02}$      \\ 
v$_{LOS}^{3}$ [km s$^{-1}$]  & -                 & -                 & -                 & 38.4$^{+14.1}_{-6.1}$    & 39.8$^{+4.7}_{-4.5}$    & 38.6$^{+4.0}_{-3.8}$      \\ 
w$_{T}^{3}$ [km s$^{-1}$]    & -                 & -                 & -                 & 54.03$^{+15.82}_{-8.95}$    & 57.84$^{+3.96}_{-3.89}$    & 51.75$^{+3.59}_{-3.57}$      \\ 
$\beta^{3}$                  & -                 & -                 & -                 & [$=\beta^{5}$]    & [$=\beta^{5}$]    & [$=\beta^{5}$]      \\ 
\hline
B$_{z}^{4}$ [G]              & -                 & -                 & -                 & [$=B_{z}^{5}$]    & [$=B_{z}^{5}$]    & [$=B_{z}^{5}$]      \\ 
$\Delta\tau^{4}_{red}$       & -                 & -                 & -                 & 0.87$^{+0.43}_{-0.37}$    & 0.91$^{+0.12}_{-0.12}$    & 0.86$^{+0.12}_{-0.12}$      \\ 
v$_{LOS}^{4}$ [km s$^{-1}$]  & -                 & -                 & -                 & 8.7$^{+7.4}_{-1.4}$    & 8.0$^{+1.2}_{-1.1}$    & 9.3$^{+2.0}_{-1.3}$      \\ 
w$_{T}^{4}$ [km s$^{-1}$]    & -                 & -                 & -                 & 15.12$^{+3.47}_{-4.46}$    & 15.31$^{+0.98}_{-0.91}$    & 14.03$^{+1.27}_{-1.39}$      \\ 
$\beta^{4}$                  & -                 & -                 & -                 & [$=\beta^{5}$]    & [$=\beta^{5}$]    & [$=\beta^{5}$]      \\ 
\hline
B$_{z}^{5}$ [G]              & -                 & -                 & -                 & -2381$^{+184}_{-78}$    & -2207$^{+62}_{-63}$    & -2607$^{+60}_{-60}$      \\ 
$\Delta\tau^{5}_{red}$       & -                 & -                 & -                 & 7.92$^{+0.96}_{-1.88}$    & 6.74$^{+0.39}_{-0.37}$    & 10.33$^{+0.72}_{-0.63}$      \\ 
v$_{LOS}^{5}$ [km s$^{-1}$]  & -                 & -                 & -                 & 1.0$^{+0.5}_{-0.3}$    & 1.0$^{+0.1}_{-0.1}$    & 0.9$^{+0.1}_{-0.1}$      \\ 
w$_{T}^{5}$ [km s$^{-1}$]    & -                 & -                 & -                 & 6.95$^{+0.73}_{-0.49}$    & 7.21$^{+0.23}_{-0.16}$    & 6.3$^{+0.22}_{-0.21}$      \\ 
$\beta^{5}$                  & -                 & -                 & -                 & 3.69$^{+0.09}_{-0.04}$    & 3.49$^{+0.02}_{-0.02}$    & 3.05$^{+0.02}_{-0.02}$      \\ 
\hline
\enddata
\tablenotetext{a}{The height is estimated from the feature morphology and previous stereoscopic analysis.  See text for details.}
\end{deluxetable*}

\begin{figure*}
    \centering
    \includegraphics[width=0.95\textwidth]{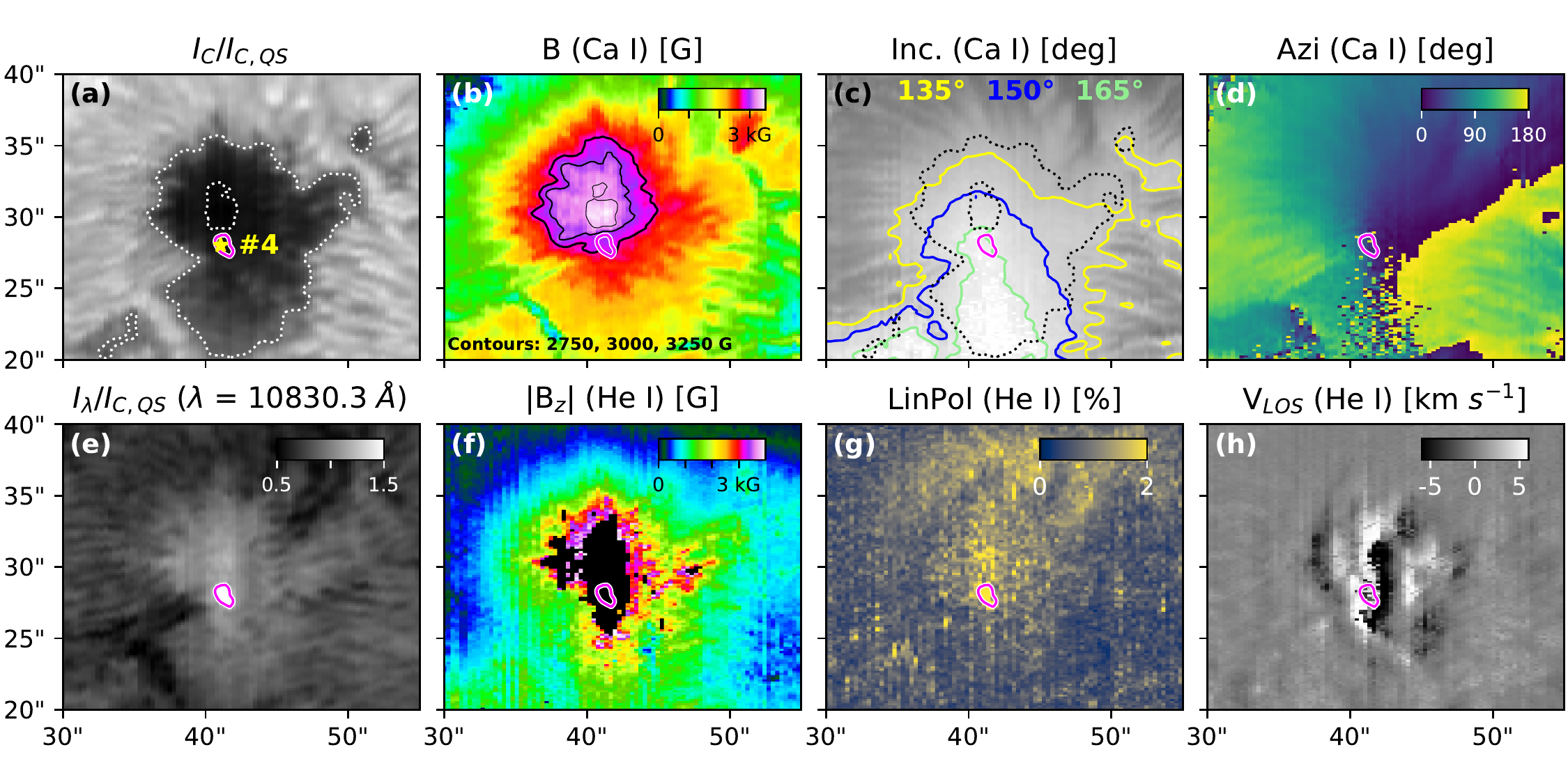} 
    \caption{Maps of the photospheric and chromospheric properties of the sunspot umbra in relation to the location of the downflow emissive kernel.  The photospheric magnetic field vector (panels b, c, and d) is inferred using a Milne-Eddington inversion of the \ion{Ca}{1} 10839 \AA{}line.  The longitudinal magnetic field strength ($B_z$) in panel (f) and Doppler velocities in panel (h) for helium result from constant-property slab inversions of the \ion{He}{1} 10830 \AA{} triplet. Negative velocities correspond to blue-shifted material. Magnetic field angles reference the line-of-sight with the azimuthal reference parallel to the solar equator. Also shown in panel (g) is the percent of \ion{He}{1} linear polarization signal.}
    \label{fig:photo_ME}
\end{figure*}

\subsection{Photospheric and Chromospheric Magnetic Field Maps}

In Figure~\ref{fig:photo_ME} we extend our analysis to the photospheric and chromospheric properties of the entire sunspot, in order to place the downflow properties in context to the surrounding area.  The vector magnetic field of the photosphere is inferred via inversions of the deconvolved full Stokes data, using the shallow \ion{Ca}{1} 10839 \AA{} line formed in the deep photosphere, and the Unno-Rachkovsky solutions for a Milne Eddington atmosphere \citep{landidegl2004}.  In the figure, we report the total field amplitude (B), the inclination along the line-of-sight direction, and the azimuth angle relative to the linear polarization reference direction, which is oriented parallel to the solar equator.  For validation, we did a cursory comparison between the retrieved umbral field strengths with that reported in Level 2 Hinode Spectropolarimeter \citep{kosugi2007} data on this date for a scan starting at 7:27 UT, which favorably showed values as high as ${\sim}$3150 G, comparable to those shown in panel (b) of Figure~\ref{fig:photo_ME}. 

The \ion{He}{1} 10830 \AA{} Stokes spectra across the sunspot are inverted using a single constant-property slab model.  Due to the level of noise and weak linear polarization signals (panel g), the inclination and azimuth of the field vector are not well recovered across the sunspot.  Figure~\ref{fig:photo_ME} displays instead the absolute value of the longitudinal magnetic field intensity ($|B_{z}|$).  Furthermore, due to the aforementioned presence of complex and emissive \ion{He}{1} profiles in the dark umbral core, we do not have high confidence that the magnetic field values returned in the affected areas are reliable.  We color these areas black in the $|B_{z}|$ map by masking those areas where an emission signal larger than 4\% is observed in the corrected data. We find that these blacked out regions lie along two vertical stripes near X = 39\arcsec and 42\arcsec. Although the fits in these regions are not satisfactory, we do find in the map of Doppler velocity (v$_{LOS}$) evidence that these regions correspond to sharp spatial-temporal transitions, from redshifts to blueshifts, consistent with umbral flashes. 


\section{EUV Data Analysis}

\begin{figure}
    \centering
    \includegraphics[width=0.4\textwidth]{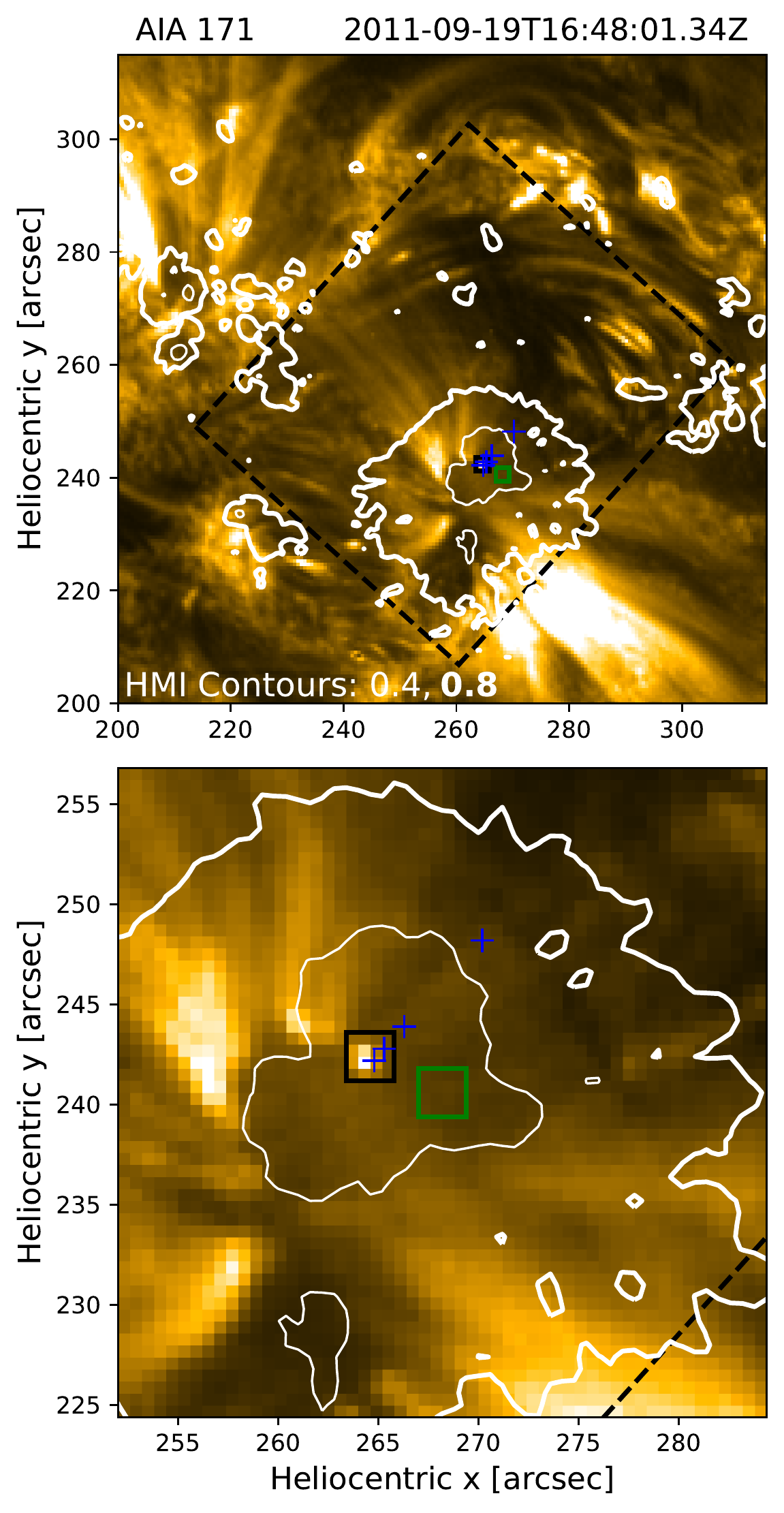}
    \caption{A context AIA 171 \AA{} image (top panel) with the FIRS scanned field of view shown by the dashed black line. Also shown in solid white lines are contours of the HMI continuum (not PSF corrected) at 0.4 and 0.8 normalized to the local quiet sun. The lower panel shows an enlarged region around the active region umbra. The blue plus signs indicate the location of the FIRS pixels discussed in the preceding sections. The black and green boxes indicate the impact and quiet umbral regions selected for emission measure analysis. The animated version of this figure shows context AIA 171 images covering the full timeseries analyzed with a label active while the FIRS scanned the umbral part of the active region.}
    \label{fig:aia_context}
\end{figure}

The abrupt termination of supersonic velocities at the location of enhanced subsonic emission in the neutral helium spectra are highly suggestive of the downflowing material undergoing a standing shock transition. To extend our thermal coverage, we investigate time series of SDO/AIA EUV imaging data coaligned with the FIRS data (see Figure~\ref{fig:aia_context}). 

\subsection{Light Curves}

The FIRS observations indicate the downflow terminus is $\approx$1\arcsec wide. For the AIA analysis we selected a square area 2.4 x 2.4 arcsec$^2$ (4 x 4 pixels) that overlaps this region (black square in Figure~\ref{fig:aia_context}). Analogously we selected a quiet nearby umbral region of similar size, shown in green, to act as a quiet-background umbra region. While the selection of the quiet region is arbitrary, analysis of several such nearby regions generally showed similar behavior.

\begin{figure}
    \centering
    \includegraphics[width=0.45\textwidth]{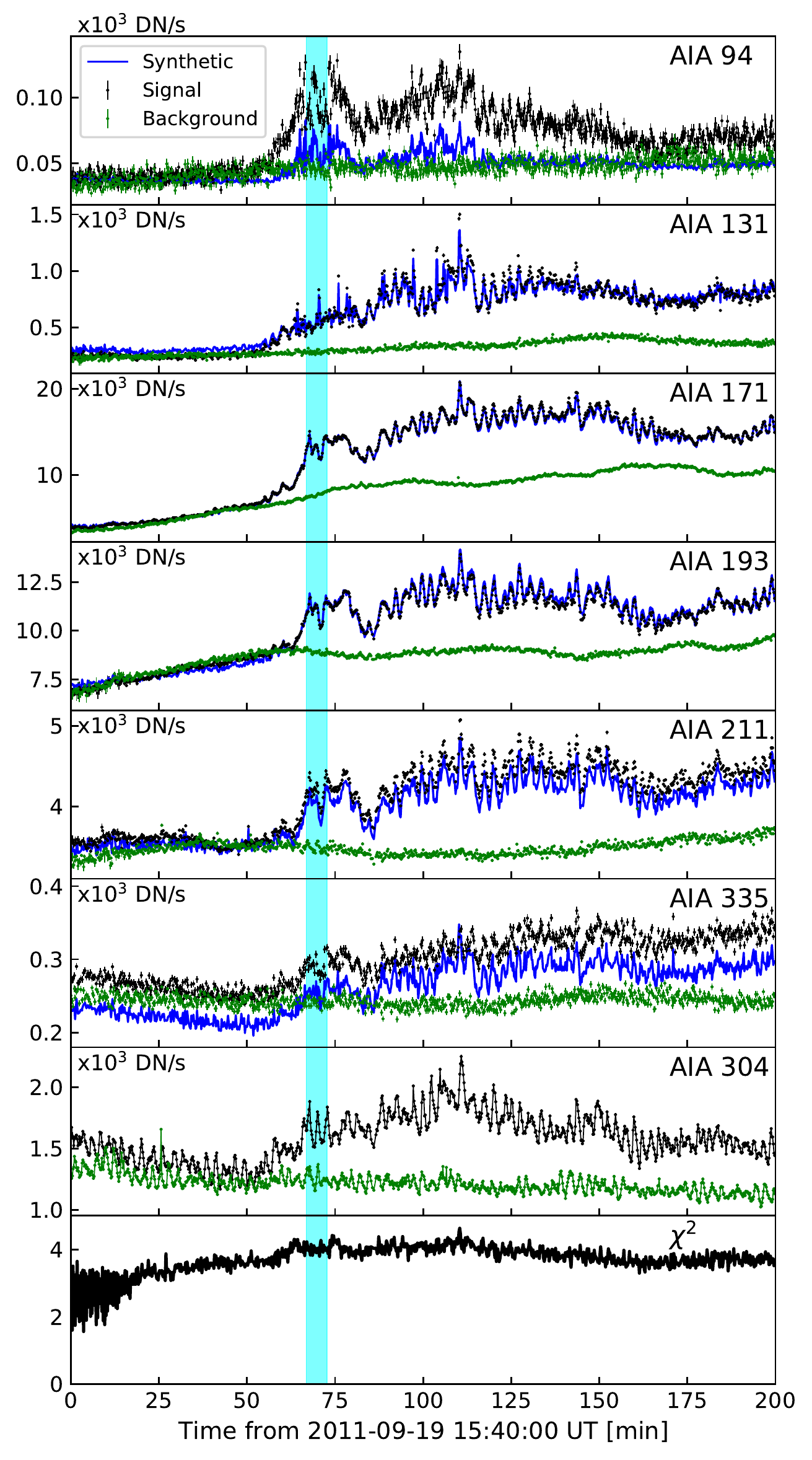}
    \caption{Timeseries of the integrated intensity with measurement errors inside the impact `signal' region and quiet umbral background marked by the black and green boxes in Fig.~\ref{fig:aia_context}. The solid blue lines show synthetic AIA signals computed from the EM(T) inversion ($\chi^2$ in bottom panel) for the impact region.  The cyan strip indicates the time range when the FIRS slit crosses the umbra.}
    \label{fig:aia_light_curves}
\end{figure}

In Figure~\ref{fig:aia_light_curves} we show the temporal variation of the integrated signal in the impact and quiet background umbra. All the filters show a sharp increase in brightness within the `signal' region (black square in Fig.~\ref{fig:aia_context}) starting $\approx$15 min before the FIRS slit position begins to cross the umbra.  This may be an indication that downflowing material had already started impacting the umbra before the FIRS scan started and continued for at least one hour after the end of the FIRS scan. The light curves in the signal area also show periodic fluctuations, but only after the initial intensity increase, with a period of ${\sim}$2.5 min, a frequency similar to chromospheric and transition region umbral oscillations and suggestive that the EUV increased emission originates below coronal heights.

\subsection{Differential Emission Measure Inversions}

Emission measure (EM) analysis is a useful tool for studying changes in the thermal state of the plasma. The column EM is defined as:
\begin{equation}
    EM(T) = \int_{0}^{\infty} n^2_e(T,z) dz \ (cm^{-5}).  \label{eqn:em}
\end{equation}
where $n_e$ is the electron density at temperature T and coordinate z along the line of sight. Using the umbral data from six of the AIA filters (94, 131, 171, 193, 211 and 335 \AA{}), we applied the inversion method described by \citet{cheung2015} to analyze temporal changes in the EM(T) and radiative cooling rates.  The AIA 304 \AA{} channel, which shows a comparable response to the downflow, is not used for the inversion as it is not adequately modeled as optically thin emission in the coronal approximation. 

The AIA temperature response functions were generated using the \textit{ch\_aia\_resp} function included in CHIANTI v9.0.1 \citep{dere2019} using the default ionization fractions and a constant pressure of p/$k_{B}$ = $10^{15}$ K cm$^{-3}$, where $k_{B}$ is the Boltzmann constant. Variations in the assumed elemental abundances primarily impact the amplitude of the calculated EM(T) but the shape and temporal variation patterns qualitatively remain the same. The EMs shown here use photospheric abundances from \citet{asplund2009}. 

For the inversion we used a logarithmic temperature grid spanning log(T[K]) $\in$ [5.5, 7.4] with 20 equally spaced bins of width $\Delta\log$(T) = 0.1. The inversion code devised by \citet{cheung2015} accepts as a starting condition sets of truncated Gaussian basis functions of varying widths centered on each temperature bin (see Figure 15 in \citet{cheung2015} for a visual representation of one such basis matrix). In searching for a possible solution for EM(T), the algorithm imposes a sparsity constraint that minimizes the number of basis functions from this starting set that are used to calculate the best fit solution. Tests done varying the number and widths of input basis functions show small improvements in the $\chi^2$ for the fits are possible, but the resulting qualitative behavior of EM(T) over time does not change. For the results presented here we settled on using three sets of Gaussian basis functions with varying widths $\sigma = \{0.0, 0.1, 0.2\}$ in log(T) units. 

\begin{figure*}
    \centering
    \includegraphics[width=0.95\textwidth]{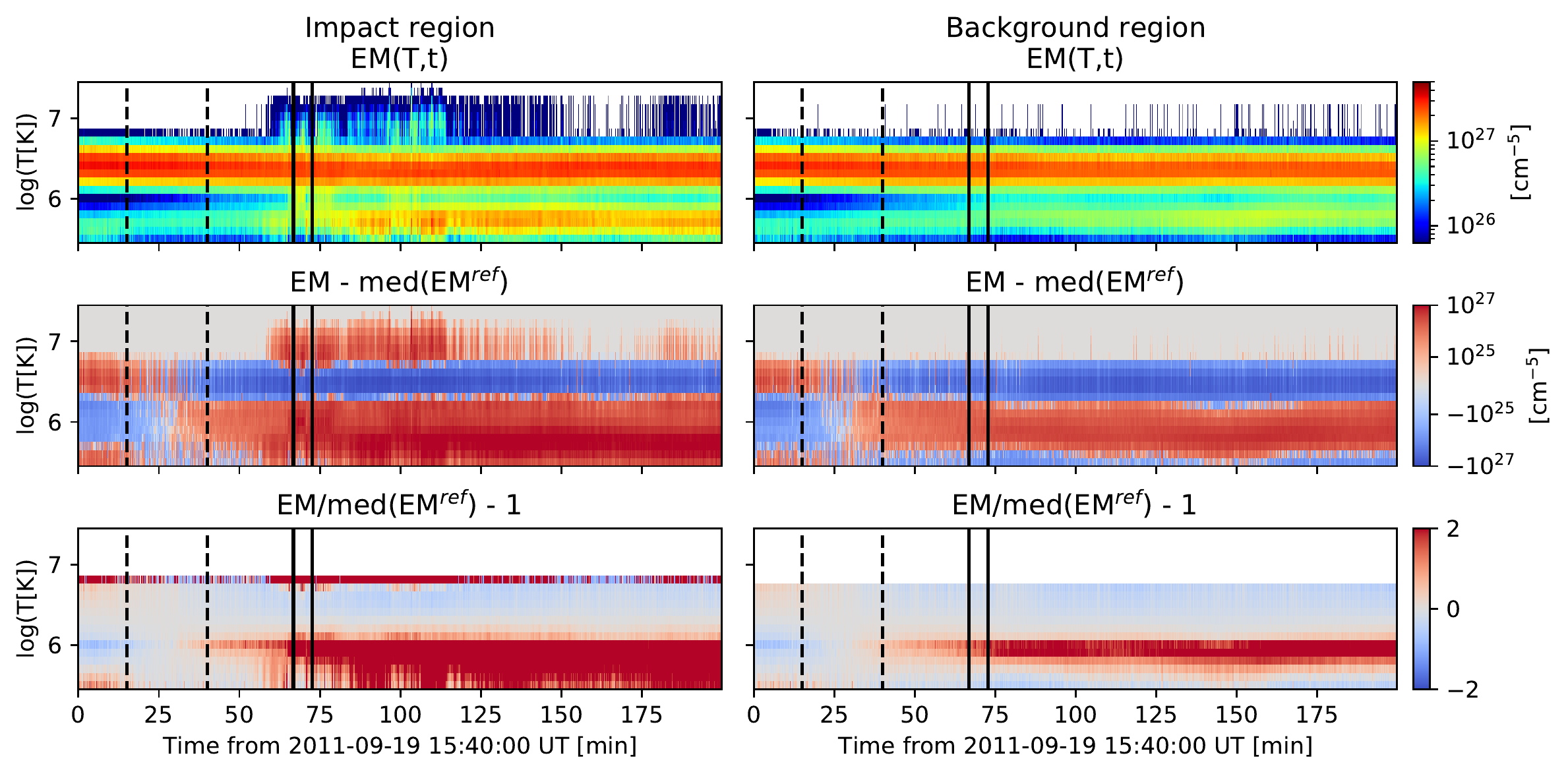}
    \caption{Timeseries of the average inverted EM(T) for the impact region (left) and the quiet umbra region (right). The dashed black lines delineate an initial quiet pre-downflow period for which we calculate a median EM(T) and subtract from subsequent timesteps to assess the relative changes in the emission measure. The solid black lines indicate the FIRS observation time window for the active region umbra.}
    \label{fig:dem_analysis}
\end{figure*}

Figure~\ref{fig:dem_analysis} shows the timeseries of inverted EMs for the impact and quiet umbra regions. The inversions are calculated for each AIA pixel independently. The results show the average EM(T) values over the 16 pixels included in the impact and quiet umbra regions selected in the preceding section. While both regions show an increase in the EM(T) at temperatures below 1 MK, the impact region shows larger absolute and relative increases compared to the time period before the downflow commenced.  The signal region also shows brief increases in EM(T) for temperatures between $10^{6.8}$ and $10^{7.3}$ K.

\subsection{Radiative losses}

From the EM(T) timeseries it is possible to calculate the total power radiated by the plasma in the inverted temperature range. The radiated power as a function of time t can be written as \citep[see, \textit{e.g.},][]{aschwanden2005}:
\begin{equation}
    P_{rad}(t) = \int_{T_1}^{T_2} EM(T,t) \Lambda(T) AdT \ (erg \ s^{-1}).
\end{equation}
where A is the area of the emitting region (measured in cm$^2$) and the radiative loss function $\Lambda(T)$ (measured in erg cm$^3$ $s^{-1}$) is primarily a function of temperature with a smaller dependence on density. We calculate $\Lambda(T)$ using CHIANTI and the same elemental abundance as for the EM(T) inversions.  For electron densities between $10^{6}$ and $10^{12}$ cm$^{-3}$ we find a variation of $<10\%$ over the entire temperature range used for the DEM inversion.  

\begin{figure*}
    \centering
    \includegraphics[width=0.85\textwidth]{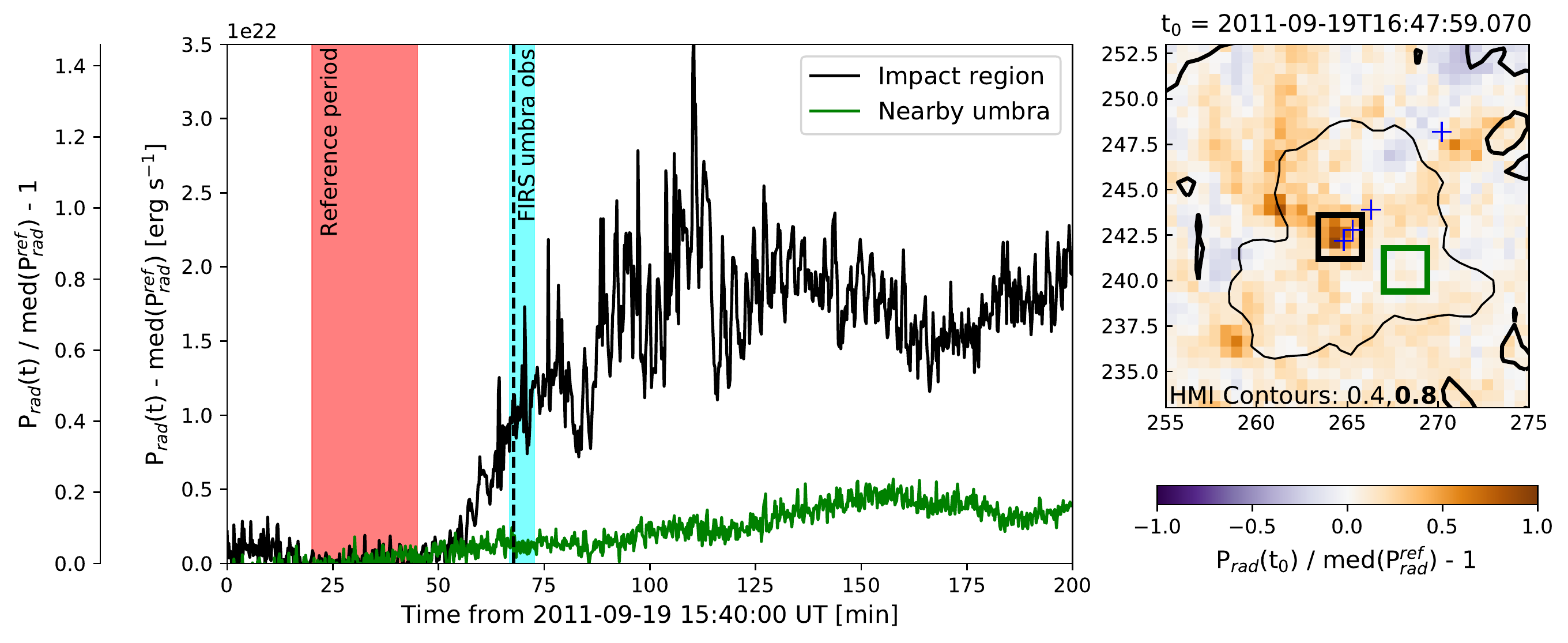}
    \caption{The left panel shows timeseries of excess radiated power relative to the median radiated power over the reference period delineated by the red box. The excess is given in absolute and relative units. The right panel shows the excess relative enhancement in power at a timestep t$_0$ when FIRS was scanning the umbra (marked by dashed line in right panel), while the animation available online shows the full timeseries. The black and green squares indicate the impact and quiet sun pixels summed to calculate the timeseries shown in the left panel. The blue markers indicate the locations of the FIRS pixels referenced in the analysis. \textit{(An animation of this figure is available)}}
    \label{fig:cooling_rates}
\end{figure*}

Figure~\ref{fig:cooling_rates} shows the excess power radiated in the two selected regions relative to a time before the downflow event begins. The impact region shows significant enhancements in radiated power throughout the downflow event, with periods when the radiative rate is nearly doubled relative to the initial state. This is in contrast to the nearby umbra which only show modest enhancements ($<20\%$) relative to the initial state. The radiative losses also exhibit the temporal oscillations with a period of ${\sim}$2.5 min. 

\begin{figure}
    \centering
    \includegraphics[width=0.45\textwidth]{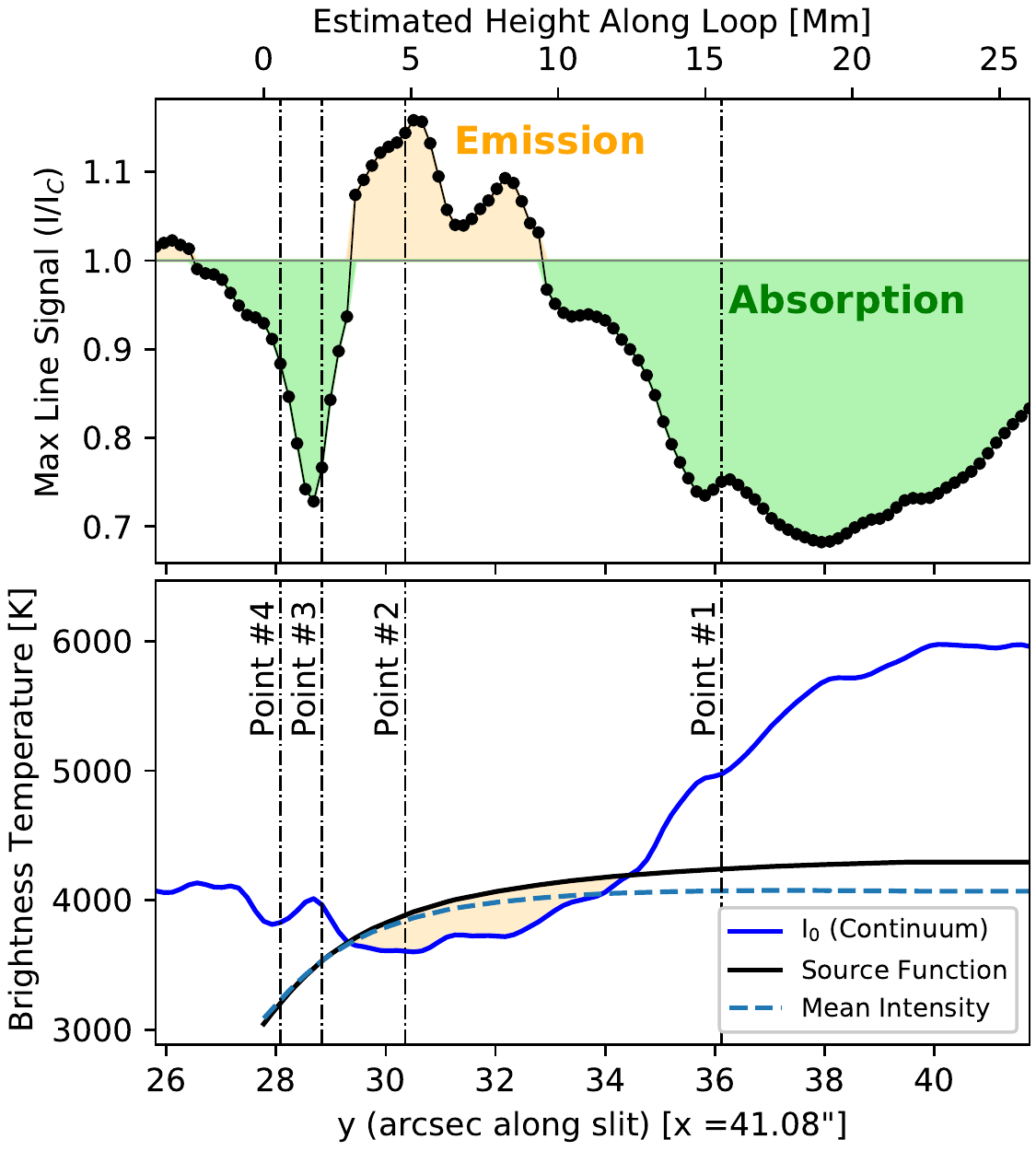}
    \caption{(top) Peak line signal normalized to the local continuum for the redshifted satellite component of \ion{He}{1} observed between 10835.38 and  10837.7 \AA{} and as a function of the y-axis coordinate along the slit at x = $41.08\arcsec$ (see Fig.~\ref{fig:image_spec}). (bottom) Comparison between the background continuum (in temperature units) and the calculated source function, for the collision-free radiative-equilibrium case ($\beta$=1), where the mean intensity of the exciting radiation field at 10830 \AA{} is derived from the observations.}
    \label{fig:emiss_abs}
\end{figure}

\begin{figure}
    \centering
    \includegraphics[width=0.45\textwidth]{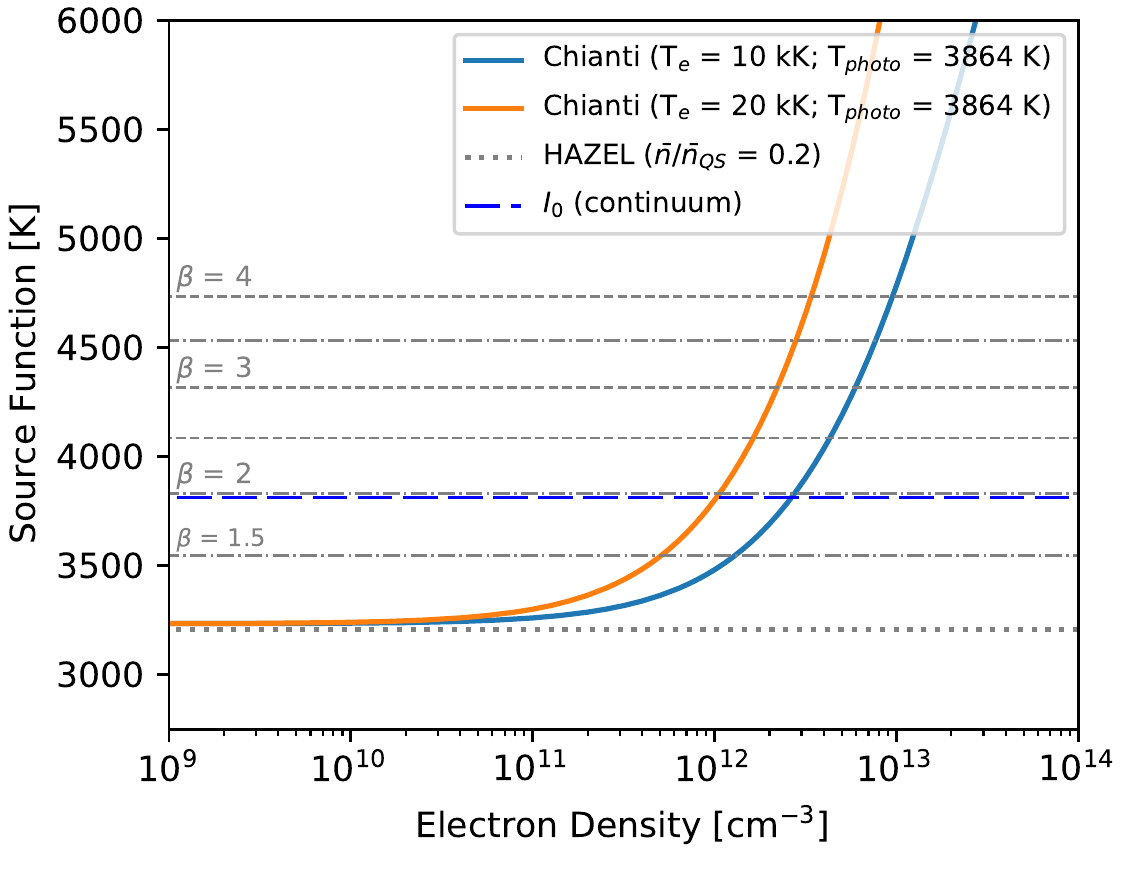} 
    \caption{Source function calculations (in temperature units) as a function of electron density and temperature for the  \ion{He}{1} 10830.3398 \AA{} transition ($2p ^{3}P_{2} - 2s ^{3}S_{1}$) using {\sc Chianti} v9.0 and assuming the radiative excitation conditions for point \#4.  $\beta$ is the source function enhancement scalar relative to the {\sc Hazel} source function, which is shown as a constant value for all densities as {\sc Hazel} does not include the collisions.}
    \label{fig:source_beta}
\end{figure}

\section{Results and Discussion}\label{sec:discussion}

We now discuss the main characteristics of the downflow and its impact with the lower atmosphere.  We further show how the observed signatures are consistent with a strong radiative shock that is uniquely observed to penetrate the lower umbral atmosphere. 

\subsection{Dynamics}\label{sec:dynamics}

The observed redshifts reach values up to 196 km s$^{-1}$ (see Table~\ref{tab:fit_values_master}).  In a statistical study, with Doppler coverage up to ${\sim}75$ km s$^{-1}$, \citet{aznar_cuadrado2007} found \ion{He}{1} SSDs extending to redshifts of 60 km s$^{-1}$ and always observed in absorption.  SSDs observed at TR temperatures by IRIS show emission components, well separated from the line core intensity, with redshifts up to ${\sim}$ 200 km s$^{-1}$ \citep{kleint2014,samanta2018}.  High-speed (${\sim}$180 km s$^{-1}$) transverse motion of raining material has been observed off-limb with IRIS observations and in \ion{He}{1}  \citep{schad2017,schad2018}.  Using the stereoscopic determined loop inclination of $23^{\circ}$ (relative to the line-of-sight), the equivalent radial-directed speeds observed here peak near 212 km s$^{-1}$.  If we assume free fall motion starting at rest, the estimated initial height of the material is ($v^{2} / 2g_{\odot} \approx$) 82 Mm, similar to the location of the highly suspended filament material associated with the source of this drainage in our earlier work (see Figure 4 of \citet{schad2016}). 

A closer look at the kinematics show a non-uniform acceleration in the lower portion of the downflow.  In Figure~\ref{fig:image_spec}, below $y=31\arcsec$, the rate of Doppler velocity change along the slit for the redshifted component increases.  We recognize the estimated heights naively assume a radial loop even at the lower heights that are not well imaged with stereoscopic methods.  Using these estimates, however, the linearized acceleration of the redshifted material between points \#1 and \#2, assuming a continuous flow and using the fitted values in Table~\ref{tab:fit_values_master}, is $0.5(v_{2}^{2} - v_{1}^2)/(h_{2} -h_{1}) = -292 \pm 17$ m s$^{-2}$, roughly in agreement with gravitational acceleration ($g_{\odot} = 274$ m s$^{-2}$).  However, the estimated acceleration increases to $-536^{+116}_{-101}$ m s$^{-2}$ between points \#2 and \#4, whose estimated height difference is ${\sim}$4.2 Mm.  Only if the height difference is grossly underestimated by $>4$ Mm would the inferred acceleration be consistent with solar gravitational acceleration.  We note that if we repeat the height estimate of the source location under the free fall assumption, using the velocities at Point \#1 at an estimated height of 15 Mm where the acceleration is consistent with free fall, then we revise our earlier estimate to 74 Mm, which remains roughly compatible with the highly suspended filament material described in our earlier work.

Increases in the flow acceleration near the footpoint are not commonly reported for TR SSDs.  \citet{kleint2014} did observe peak velocities faster than expected for free fall acceleration along a curved loop, which would require a non-zero initial velocity for the free-fall scenario.  Meanwhile, the majority of coronal rain events show acceleration less than free fall \citep[see, \textit{e.g.},][]{antolin2012sharp}.  \citet{muller2005} suggested high speeds could be created when material flows along a pre-evacuated loop, though their simulations suggest the material will still be decelerated near the flow footpoint. 

In this case, the footpoint is embedded in the sunspot and the flow persists for more than an hour based on the AIA observations.  If we approximate it as continuous, compressible field-aligned flow, increases in the velocity can be expected when the cross-sectional area (A) reduces and the magnetic field intensifies.  Estimates of the density using the \ion{He}{1} profiles are not possible when the signal is optically thin (see Section~\ref{sec:density_high_speed}); however, using the continuity equation ($nAV$ = constant, where n is the total number density) and magnetic flux conservation ($BA$ = constant), we can infer relative changes in the flow properties using the fitted values (Table~\ref{tab:fit_values_master}), where $v_{LOS}{\approx}v$, $B_{z}{\approx}B$, and T is inferred from $w_{T}$, \textit{i.e.} $T = M\sqrt{w_{T}^2 - \xi^2} / (2k_{B})$ where M is the mass of helium and $\xi$ is the instrumental broadening.  Assuming an ideal gas, the relative pressure change is equivalent to the relative density change multiplied by the relative temperature change. Based on this approach, Table~\ref{tab:continuity} provides the flow properties inferred from the redshifted material and expressed as a fraction relative to point \#1.  There is only modest evolution of the highly redshifted material-- a factor of two change in temperature and density while the pressure remains fairly constant (within the error bounds). 

\begin{deluxetable}{lcccc}
\tablecaption{Line width temperatures (top) and flow  properties (bottom) expressed as a fraction relative to point \#1 using the retrieved values for the PSF-corrected data in Table~\ref{tab:fit_values_master} and assuming continuous compressible flow and magnetic flux conservation. \label{tab:continuity}}
\tablehead{
  \colhead{Parameter} & \colhead{Point \#1} & \colhead{Point \#2} & \colhead{Point \#3} & \colhead{Point \#4}
}
\startdata
T [kK]    & 17.6$^{+1.9}_{-1.8}$    & 15.8$^{+4.1}_{-3.6}$    & 10.6$^{+1.8}_{-1.6}$    & 11.5$^{+8.9}_{-4.8}$      \\ 
\hline
Temp (T) & - & 0.9$^{+0.36}_{-0.27}$  & 0.6$^{+0.18}_{-0.14}$ & 0.65$^{+0.64}_{-0.31}$ \\ 
Area (A) & - & 0.55$^{+0.24}_{-0.17}$  & 0.54$^{+0.16}_{-0.13}$ & 0.34$^{+0.91}_{-0.16}$ \\ 
Dens (n) & - & 1.65$^{+0.39}_{-0.3}$  & 1.61$^{+0.37}_{-0.29}$ & 2.48$^{+0.52}_{-0.4}$ \\ 
Pres (P) & - & 1.47$^{+1.09}_{-0.64}$ & 0.97$^{+0.59}_{-0.36}$ & 1.62$^{+2.27}_{-0.91}$ \\ 
\enddata
\end{deluxetable}

\subsection{He I line formation of the high speed material}\label{sec:line_formation}

The redshifted \ion{He}{1} component is in absorption away from the sunspot, while near the highest speeds, and when projected over the darkest parts of the sunspot umbra, the \ion{He}{1} lines appear in emission.  We can explain the emission profiles as the consequence of the relative increase of the mean intensity of the radiation field exciting the material at higher heights above the sunspot compared to the intensity of the background continuum along the line-of-sight.  This is best shown in Figure~\ref{fig:emiss_abs}, where the line signal along the slit (from Figure~\ref{fig:image_spec}) is compared to the background continuum intensity ($I_{0}$, solid blue line), the line source function (black line) determined using Equation 9.14 of \citet{landidegl2004} and the level populations calculated by {\sc Hazel-2}, and the mean intensity from Figure~\ref{fig:rad_tensor} (dashed blue line).  All are given in units of temperature.  Recall that to fit the emission spectra for both point \#1 and \#2, no adhoc scaling of the line source function was necessary ($\beta=1$). In the optically thin case, Equation~\ref{eqn:opt_thin} shows the relative amplitude of the source function and the background intensity dictate whether the line is in emission or absorption.  Indeed, we find the line source function is slightly larger than the background intensity within the umbra, which explains the point \#2 emission.  For point \#3, the modeled profile is optically thick and required a marginal scaling of the line source function ($\beta=1.18$).  However, the line remains in absorption, which is consistent with the figure wherein the continuum brightness temperature exceeds the mean intensity at low heights.  This level of agreement provides us some confidence that our height estimates are credible to within one or two megameters. 

One consequence of this finding is that the detection of SSDs within sunspot umbrae in \ion{He}{1} 10830 \AA{} lines may be more complicated, especially for large dark sunspots, as the line will not remain in absorption and could become weak and harder to detect, especially in the presence of telluric and photospheric line absorption. 

\subsection{He I density estimates of the high speed material}\label{sec:density_high_speed}

Similar to point \#3, the highly redshifted component of point \#4 requires scaling of the modeled source function ($\beta^{1}=1.59$ in Table~\ref{tab:fit_values_master}); however, this scaling is especially sensitive to the estimated height on account of the mean intensity of the photoexcitation rapidly increasing between heights of 0 and 5 Mm. That said, an enhanced source function could also be a sign of electron collisions at higher densities. In Figure~\ref{fig:source_beta}, we evaluate the expected enhancement of the source function as a function of electron density using CHIANTI v9 calculations under the coronal approximation \citep{dere2019}.  Similar to {\sc Hazel-2}, these calculations do not incorporate all the necessary mechanisms to understand the helium triplet population; however, they do provide an initial estimate on the role of collisions on the source function, which is not included in {\sc Hazel-2}.  Our calculations assume the same degree of photoexcitation as used to model the point \#4 spectra ($\bar{n}/\bar{n}_{QS} = 0.2$), which in the formulation of CHIANTI corresponds to excitation by a radiation field of 3864 K at zero height.  The CHIANTI and {\sc Hazel} source functions for $\beta$=$1$ agree at low density.  For an electron temperature $T_{e} = 10$ kK, $\beta=1.59$ for point \#3 implies an electron density of ($\log n_{e} \approx$) 12.1.  This is on the higher end of what could be expected and is likely an upper estimate.  \citet{samanta2018} reported ($10.17 \pm 0.69$ dex) for IRIS observed SSDs at transition region temperatures.  \citet{antolin2015} estimated core coronal rain densities, inferred using EUV absorption due to hydroden and helium, to be 10.3 to 11.4 dex. \citet{scullion2016} estimated 11.9 to 12 dex for flare-driven coronal rain. 

To gain a better understanding of how the estimated heights affects the inferred $\beta$ values for point \#4, we repeated the above analysis, including the model fitting, by assuming the height is 1 Mm higher than our nominal estimates.  In this case, the height is taken to be 1.57 Mm and $\bar{n}/\bar{n}_{QS} = 0.43$ and the inferred values of $\beta^{1}$ (corresponding to the high speed absorption) and $\beta^{5}$ (corresponding to the emission) change to 1.17 and 2.73, respectively.  The former would imply an electron density of ${\sim}$11.3 dex [cm$^{-3}$] in the high speed component, which is more in line with previous observations.

Oddly, the second, less-redshifted, component in the point \#3 profile required a very small source function, $\beta \approx 0.1$, to be fit. We do not have a good explanation for this small value; though, we discuss below that non-equilibrium effects may be important.  It is also possible that it is an artifact of our fitting methodology.  In order to preserve the relative shape that the optically thick faster component imparts on the profile, the broad blue component is forced to take on a small optical thickness, which then drives the reduction in the source function in order to account for the total absorption.  Alternative models might be considered; however, the photospheric line blends pose a severe challenge to achieving a robust fit for a more complex model.  

\subsection{He I line formation of the near rest emission}\label{sec:density_emission}

Finally, we focus on the formation of the large emission component and its red tail in point \#4.  Three model slabs adequately reproduce the profile shape, with components at 1, 8.7 and 38.4 km s$^{-1}$.  Respectively, the optical thickness declines from a thick near-rest component ($\Delta\tau_{red} = 7.92$) to two thin components.  We expect the redward tail to be generated by large gradients in the atmosphere along the line-of-sight, which is further corroborated by the lack of a defined Stokes V redshifted component in the line wing.  In other words, there are no distinct velocity components within the tail. Note that the large line width of the fastest emissive slab (54.03 km s$^{-1}$) represents unphysical temperatures for neutral helium (\text{i.e.} 700 kK). Meanwhile, the inferred value of $\beta$ is 3.69, which is reduced to 3.05 in the event the spectral straylight has an absorption signal at the lower bound of what is expected (profile B in Figure~\ref{fig:emiss_fits_wstraylight}).  In Figure~\ref{fig:source_beta}, the corresponding electron densities are in the range of 12.8 to 13 dex.  As above, the height estimates will have an influence on the density estimate; however, this uncertainly is unlikely to lower the estimated density more than 0.5 dex given the strength of the emission.  This is also corroborated by our calculations that increase the height by 1 Mm from the nominal estimates. 

\subsection{Comparison to an isothermal radiative shock model}

The characteristics of the terminus of the observed supersonic downflow are consistent with a nearly isothermal radiative shock, wherein the shocked gas downstream of an initial normal shock radiatively cools rapidly.  First consider that in the limit of a very strong shock, the Rankine-Hugoniot jump conditions \citep[see, \textit{e.g.}, ][]{shu1992physics} permit at maximum a factor of 4 enhancement in the gas density, and a likewise factor of 4 reduction in the velocity immediately behind the shock.  For the observed flow, assuming the local sound speed is between 10 and 20 km sec$^{-1}$ at the shock, the upstream Mach number (M$_{1}$) is ${\sim}$ 20 to 10.  The expected downstream flow velocity is then ${\sim}$50 km s$^{-1}$, and the downstream temperature would increase by a factor between 30 and 125, assuming a perfect gas.  For an upstream temperature of 10 to 20 kK, as inferred from the measurements, the gas temperature would rise to between 0.3 and 2.5 MK, or 5.5 to 6.4 dex [K]. 

In an optically thin radiative shock that efficiently cools downstream of the initial shock layer \citep[see, \textit{e.g.}, ][]{shu1992physics}, the density and pressure can continue to be enhanced beyond the strong shock factor of 4 as the flow continues to decelerate.  In the limit where the final temperature (T$_{3}$) is equal to the original upstream temperature (T$_{1}$), which is approximately the case in our observations, the isothermal shock conditions give $\rho_{3} / \rho_{1} = u_{1} / u_{3} = M_{1}^{2}$, or between 400 and 100 assuming a constant local sound speed between 10 and 20 km sec$^{-1}$.  Although the density estimates we can glean from the \ion{He}{1} are approximate, this two order of magnitude density increase is in line with the estimates given above for the upstream and downstream gas. Furthermore, the Doppler extent of the helium line profile at point \#4 can be explained by the isothermal shock conditions where the ratio of the upstream and radiatively-cooled gas velocity is ${\sim}$200/1.

\subsection{Energetics}

The above scenario is also consistent with the increased emission measure inferred from the EM inversions of the SDO/AIA data.  Enhanced emission below 10$^{6}$ K in Figure~\ref{fig:dem_analysis}, peaking towards the lower temperatures of the thermal bandpass of the EM distribution, is consistent with the increased radiative cooling of the shocked material, with the radiative loss rate $\Lambda(T)$ being the highest in the range of 0.9 to $3 \times 10^{5}$ K as calculated using CHIANTI. The EM distributions also show in the top panel of Figure~\ref{fig:dem_analysis} brief heating episodes up to temperatures of 10$^{7}$ K, particularly in the time period 60 to 115 minutes after 15:40 UT, and a weak decrease of the EM between 6.3 and 6.6 dex K compared to the reference time period.  The latter is likely influenced by the evolution of the region as a whole along the line-of-sight. 

We have derived the total radiative loss rate associated with the downflow in Figure~\ref{fig:cooling_rates}, which shows values in the range of 0.5 to $2.5 \times 10^{22}$ erg s$^{-1}$.  We can compare this to the total power in the downflow.  Assuming a continuous flow, the total power can be written as
\begin{equation}
    Power = \frac{1}{2}\rho v^{3} A 
          \approx  \frac{1}{2}m_{p}n_{e} \left ( 
          \frac{1+4f_{He/H}}{1+2f_{He/H}} 
          \right )
          v^{3} A 
\end{equation} 
where $A$ is the cross-sectional area of the flow and $f_{He/H}$ is the fractional abundance of helium relative to hydrogen, assuming a H-He gas mixture.  Adopting $f_{He/H} \sim 0.08$, and taking $n_{e} \approx 10^{11}$ cm$^{-3}$, $ v = 195$ km s$^{-1}$, and A of 1 arcsec$^2$ as representative of the observed flow, the estimated downflow power is ${\sim}3.5 \times 10^{24}$ erg s$^{-1}$.  And so the downflow power is sufficient for driving the observed radiative losses. 

The inferred AIA emission measure distribution also provides some additional constraints for the electron density in the shock-heated and radiatively cooling gas.  Taking the peak columnar EM near $10^{5.8}$ K to be ${\sim}$1 to $4 \times 10^{27}$ cm$^{-5}$ from Figure~\ref{fig:dem_analysis} and attributing it to the cooling post-shocked gas radiatively over some column length, we can estimate the electron density using Equation~\ref{eqn:em} further assuming it is constant.  The length of the radiating layer, however, is not well known and the density is expected to strongly increase as the material cools.  For a large range of possible radiating lengths between 100 km to 2 Mm, we calculate densities of 11.8 to 10.8 dex $cm^{-3}$, which are compatible with the densities inferred using the helium profiles.

\subsection{Mass supply}

Similar to \citet{straus2015} and \citet{chitta2016}, the observed duration of the downflow is long compared the estimated timescale to drain the coronal loop's volume without addition mass supply.  The primary drainage channel, shown in panel (c) of Figure~\ref{fig:overview}, persists long enough for FIRS to resolve material along its entire length during the associated 23 minute scan time between the beginning of the scan to when the slit reaches the location of the downflow terminus in the umbra.  The SDO/AIA light curves in Figure~\ref{fig:aia_light_curves} indicate the increased radiative losses in the umbra near this downflow's footpoint begin just prior to the beginning of the FIRS scan and continue for ${\sim}100$ minutes.  Following \citet{chitta2016}, the drainage timescale can be estimated as $\tau_{drainage} \approx 0.1L/v$.  Assuming a semicircular coronal loop with constant cross-sectional area and a height of $h = 80$ Mm, and $v = 200$ km s$^{-1}$, $\tau_{drainage} \approx 10$ minutes.  This suggests the observed downflow needs an additional mass supply.  However, given its association with the filament material that resides above the active region complex, as described in \citet{schad2016}, and its near free fall acceleration at heights of ${\sim}15$ Mm, interpreting this flow as a siphon flow does not seem justified.

\subsection{Interpreting the He I emission profile tails}

The presence of the long redward emissive tail for point \#4 can be explained as a natural consequence of post-shock cooling during deceleration provided there is sufficient opacity in the neutral triplet.  In the simplified radiative shock picture above, assuming a strong shock, the velocity immediately behind the shock front is (${\sim}200/4 =$) 50 km s$^{-1}$; though, the observations show emission up to ${\sim}$ 140 km $^{-1}$.  This may imply that sufficient neutral helium remains present through the viscous layer of the shock in order to have emission at high velocities.  Furthermore, there is the odd enhanced absorption feature for point \#3 that is not easily explained. \citet{golding2014} modeled the helium triplet lines for a weaker shock, not in umbral conditions, but including the effects of non-equilibrium ionization, concluding that non-equilibrium relaxation timescales can be long compared to the dynamics and are thus important to consider.  Brief enhancements in \ion{He}{1} 10830 \AA{} absorption followed by strong emission has been observed during solar flares \citep{xu2016} and has been found in numerical calculations that include electron beam heating by \citet{ding2005},  \citet{huang2020}, and \citet{kerr2021}.  We suspect a similar scenario may be possible for enhanced collisions at the shock front of a strong downflow, as in the event studied here, but further investigations are necessary to confirm this possibility. 

\subsection{Magnetic fields: how deep does the flow penetrate?}

Finally, we consider more closely the inferred magnetic field values. The longitudinal magnetic field ($B_{z}$) inferred from the redshifted component increases from ${\sim}$461 G, at a height of ${\approx}$15.5 Mm, to ${\sim}$1351 G at or directly above the shock.  In the strong post-shock emission, $B_{z}{\sim}$2381 G within the deconvolved data set, or between ${\sim}$2207 and ${\sim}$2607 G using alternative straylight correction (Section~\ref{sec:fits_non_deconv}). The photospheric \ion{Ca}{1} magnetic field strength along the line-of-sight is ${\sim}$2800 G.  Meanwhile, the inferred longitudinal field strengths in the \ion{He}{1} line outside of point \#4, shown in Figure~\ref{fig:photo_ME}, give values of ${\sim}$ 2000 in the outer umbra and suggest field strengths in the nominal chromosphere are above this within the region probed by the downflow; however, unfortunately, it this region, the complex profiles in the PSF-corrected data do not permit suitable estimates. 

The magnetic field strength of 2.4 kG, or even the less likely lower estimate of 2.2 kG, inferred in the bright \ion{He}{1} 10830 \AA{} emission may be the largest reported to date for this line. \citet{schad2015} and \citet{joshi2017} reported peak \ion{He}{1} umbral field strengths ranging between 1.5 and 1.8 kG.  For the same data set, using the nearby \ion{Si}{1} and \ion{Ca}{1} lines, \citeauthor{joshi2017} inferred photospheric umbral field strengths between 2.5 to 2.8 kG. In a C class flare, \citet{anan2018} inferred a value of 1400 G in \ion{He}{1} 10830 \AA{}.  Meanwhile, in the He D3 5876 \AA{} line, \citet{libbrecht2019} reported values as high as 2.5 kG for a C flare above a pore, although, they remark their observed spectra can also be fit using substantially larger values (up to 5 kG). 

The uncertainties involved in measuring the magnetic field in the umbral chromosphere makes it challenging to relate the post-shock downflow field strength to the nominal (yet dynamic) chromosphere.  From our work, we have learned that correcting for straylight is essential and may effect previous analyses.  Where the neutral helium is formed in the umbra is ill-determined and will influence measurements of the height gradient of the sunspot magnetic field, which is still an open question as discussed by \citet{balthasar2018}.  
The post-shock helium emission field strength of ${\sim}$ 2.4 kG is only 400 G weaker than the underlying photosphere (2.8 kG).  Likely the difference is even less as the former value refers to only the longitudinal component whereas the latter is the total field strength.  It should also be noted that the \ion{He}{1} material is optically thick where this field is measured, and so the flow likely penetrates further than what we can estimate using this line.  If we adopt a shallow estimate for the height gradient of -0.5 G km$^{-1}$ as reviewed in \citet{balthasar2018}, then the post-shock material would be at heights less than 800 km, in the low chromosphere for most umbral models \citep[see, \textit{e.g.},][]{loukitcheva2017}.  We have not found any signatures of the downflow in the photosphere.  Ordinarily, the population of triplet state helium in the photosphere and low chromosphere is minute \citep{centeno2008}; however, in a non-equilibrium state, longer relaxation timescales for recombination from \ion{He}{2} \citep{golding2014}, together with fast downflow speeds, might populate the triplet state in the lower atmosphere. 

We also infer a kilogauss difference in the pre-shock (1.3 kG) and post-shock field (2.4 kG) for point \#4.  The shock normal is expected to be parallel with the magnetic field direction, in which case we do not expect the shock itself to change the field intensity.  Increases in the gas pressure downstream of the shock would presumably lead to a field reduction.  We suspect the difference may be related to the height difference between the pre-shock and post-shock material, which is not well constrained by our observations as already implied by the density analysis above for the upstream density.  A height difference of at least 1 Mm between the upstream high speed material and the post-shock (lowest speed) material could be possible. 

\section{Summary}

We have reported \ion{He}{1} spectropolarimetric observations of a supersonic downflow (SSD) into a sunspot umbra that exhibits signatures consistent with a strong isothermal radiative shock impacting the lower atmosphere.  Recent studies, \textit{i.e.} \citet{samanta2018} and \citet{nelson2020b}, have begun to quantify the extent of the role of SSDs in sunspot umbral atmospheres and their evolution; though, more studies are necessary.  Our observations are unique in that measurements of the magnetic field strength in the optically thick post-shocked helium have values only a few hundred Gauss less than the underlying photosphere. This suggests that penetration of the flow well into the chromosphere, and perhaps the upper photosphere, might be possible for such strong flows.  These observations have also provided a detailed spectral and polarimetric view of the signatures of neutral helium through each phase of the shock, which are important to consider further with non-equilibrium models for strong downflow events like that observed here or during solar flares. 

\acknowledgments

We thank the referee for their careful reading of the manuscript and helpful comments.  The National Solar Observatory (NSO) is operated by the Association of Universities for Research in Astronomy, Inc. (AURA), under cooperative agreement with the National Science Foundation.  FIRS has been developed by the Institute for Astronomy at the University of Hawai‘i, jointly with the National Solar Observatory (NSO). The FIRS project was funded by the National Science Foundation Major Research Instrument program, grant number ATM-0421582. {\sc Chianti} is a collaborative project involving George Mason University, the University of Michigan (USA), University of Cambridge (UK) and NASA Goddard Space Flight Center (USA).   SDO data are the courtesy of NASA/SDO and
the AIA and HMI science teams.  This research has made use of
NASA’s Astrophysics Data System.

\bibliography{main}{}
\bibliographystyle{aasjournal}

\appendix

\section{Linear polarized spectra}\label{appendix:linpol}

\begin{figure*}
    \centering
    \includegraphics[width=0.99\textwidth]{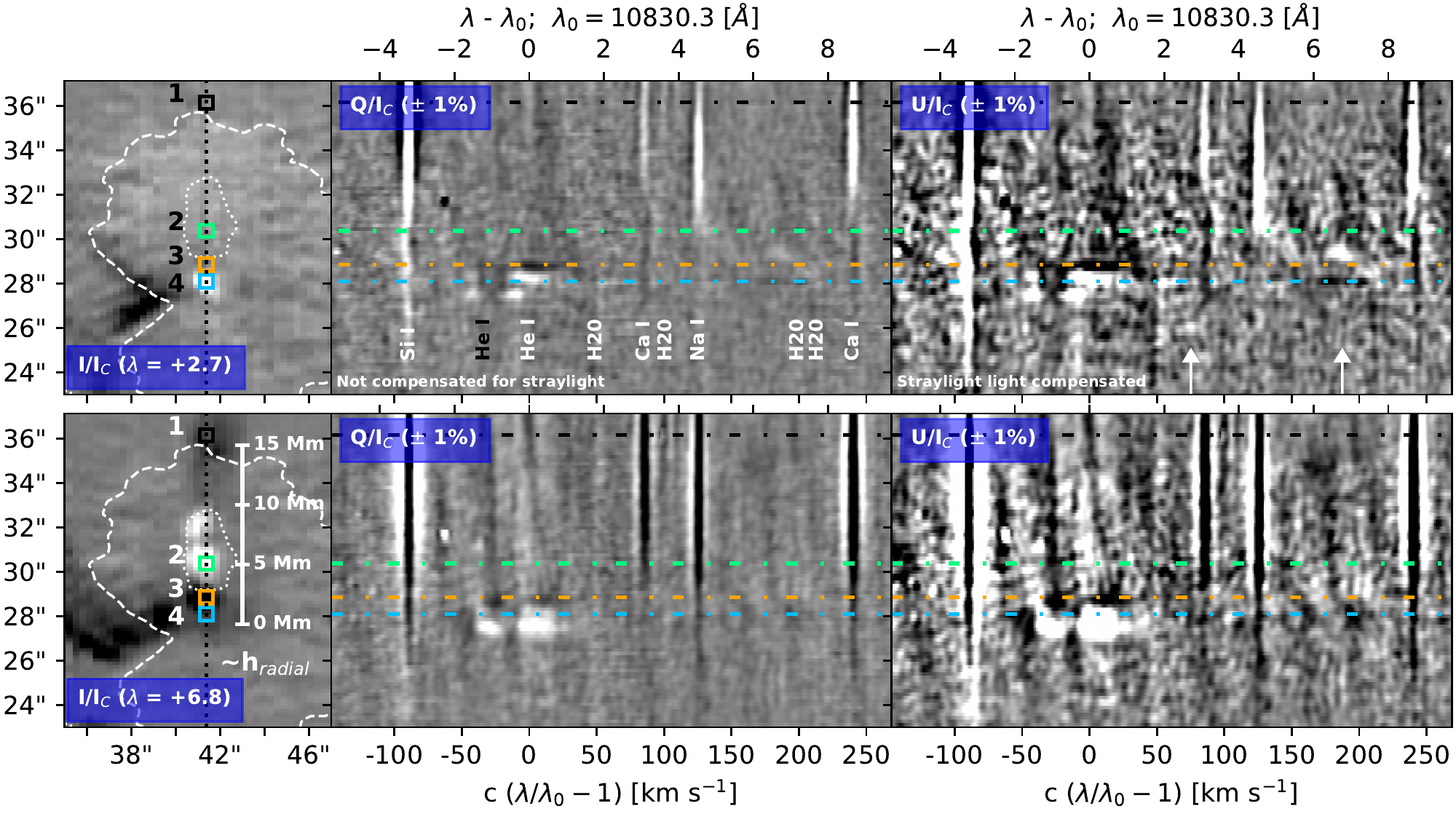}
    \caption{Same as Figure~\ref{fig:image_spec} but showing Stokes Q and U scaled between $-1\%$ and $+1\%$.}
    \label{fig:image_spec_QU}
\end{figure*}

Figure~\ref{fig:image_spec_QU} shows the Stokes Q and U image spectra along the observed supersonic downflow.  We do not analyze these signals in the main paper but instead show them here to be complete.  Small traces of the Q and U scattering polarized signals in the downflow, which were analyzed by \citet{schad2016}, can be seen near 10836.3 \AA{} for $34'' < y < 36''$. Within the umbra, residual interference fringes and increases in observational noise dominate; although, helium signals in the near-rest emissive component are observed near y = $28.1\arcsec$. 

\end{document}